\documentclass[useAMS]{mn2e}

\usepackage{amssymb,amsmath,psfig,times}
\def\gsim{ \lower .75ex \hbox{$\sim$} \llap{\raise .27ex \hbox{$>$}} }
\def\lsim{ \lower .75ex\hbox{$\sim$} \llap{\raise .27ex \hbox{$<$}} }

\def\kmps{{\rm\thinspace km \thinspace s^{-1}}}

\def\sc{Schwarzschild}
\def\beq{\begin{equation}}
\def\eeq{\end{equation}}

\def\sc{Schwarzschild}
\def\Omm{{\Omega_m}}
\def\Ommz{{\Omega_m^{\,z}}}

\def\Omk{{\Omega_k}}
\def\Oml{{\Omega_{\Lambda}}}

\title[Heaviest black holes of jetted AGNs]
{Chasing the heaviest black holes of jetted Active Galactic Nuclei}  
\author[G. Ghisellini et al.]
{G. Ghisellini$^1$\thanks{Email:
gabriele.ghisellini@brera.inaf.it}, R. Della Ceca$^2$, 
M. Volonteri$^3$,
G. Ghirlanda$^1$, 
F. Tavecchio$^1$, 
\newauthor{L. Foschini$^1$, G. Tagliaferri$^1$, F. Haardt$^{4,5}$,
G. Pareschi$^1$, J. Grindlay$^6$}
\\
\\
$^1$INAF -- Osservatorio Astronomico di Brera, Via Bianchi 46, I--23807 Merate, Italy\\
$^2$INAF -- Osservatorio Astronomico di Brera, Via Brera 28, I--20100 Milano, Italy\\
$^3$Astronomy Department, University of Michigan, Ann Arbor, MI 48109 \\
$^4$Universit\'a dell'Insubria, Dipartimento di Fisica e Matematica, Via Valleggio 11, 
I--22100 Como, Italy;\\
$^5$ INFN, Sezione di Milano-–Bicocca, I--20126 Milano, Italy\\
$^6$Harvard--Smithsonian  CfA, 60 Garden Street  Cambridge MA 02138, USA \\
}

\begin{document}  

\maketitle

\begin{abstract}
We investigate the physical properties of the 10 blazars at redshift
greater than 2 detected in the 3--years all sky survey performed by 
the Burst Alert Telescope (BAT) onboard the {\it Swift} satellite.
We find that the jets of these blazars are among the most powerful known.
Furthermore, the mass of their central black hole, inferred from the optical--UV
bump, exceeds a few billions of solar masses, with accretion luminosities 
being a large fraction of the Eddington one.
We compare their properties with those of the brightest blazars
of the 3--months survey performed by the Large Area Telescope (LAT) onboard
the {\it Fermi} satellite.
We find that the BAT blazars have more powerful
jets, more luminous accretion disks and larger black hole masses than LAT blazars.
These findings can be simply understood on the basis of the blazar sequence,
that suggests that the most powerful blazars have a spectral
energy distribution with a high energy peak at MeV (or even sub--MeV)
energies.
This implies that the most extreme blazars can be found more efficiently 
in hard X--rays, rather than in the high energy $\gamma$--ray band.
We then discuss the implications of our findings for
future missions, such as the {\it New Hard X--ray Mission (NHXM)} and especially
the {\it Energetic X--ray Imaging Survey Telescope (EXIST)} mission
which, during its planned 2 years all sky survey, is expected to detect thousands of 
blazars, with a few of them at $z\gsim$6.
\end{abstract}

\section{Introduction}

Ajello et al. (2009, hereafter A09) recently published the list 
of blazars detected in the all sky survey by the 
Burst Alert Telescope (BAT)
onboard the {\it Swift} satellite, between March 2005 and March 2008.
BAT is a coded mask designed to detect Gamma Ray Bursts (GRBs),
has a large field of view ($120^\circ\times 90^\circ$, partially coded) and
is sensitive in the [15--150 keV] energy range.
This instrument was specifically designed to detect GRBs,
but since GRBs are distributed isotropically in the sky, BAT, 
as a by--product, performed an all sky survey with a reasonably 
uniform sky coverage, at a limiting sensitivity of the order 
of 1 mCrab in the 15--55 keV range 
(equivalent to $1.27\times 10^{-11}$ erg cm$^{-2}$ s$^{-1}$) 
in 1 Ms exposure (A09).
Taking the period March 2005 -- March 2008, and evaluating the image
resulting from the superposition of all observations in this period,
BAT detected 38 blazars (A09), of which 26 are Flat Spectrum Radio Quasars
(FSRQs) ad 12 are BL Lac objects, once the Galactic plane ($|b<15^\circ|$) 
is excluded from the analysis.
A09 reported an average exposure of 4.3 Ms, and considered
the [15--55 keV] energy range, to avoid background problems at higher
energies.
The well defined sky coverage and sources selection criteria
makes the list of the found blazars a 
complete, flux limited, sample, that enabled A09 to calculate the luminosity
function and the possible cosmic evolutions of FSRQs and BL Lacs,
together with their contribution to the hard X--ray background.
A09 also stressed the fact that the detected BAT blazars at high redshift
are among the most powerful blazars and could be associated with powerful
accreting systems.

\begin{figure}
\vskip -0.6 cm \hskip -0.4 cm
\psfig{figure=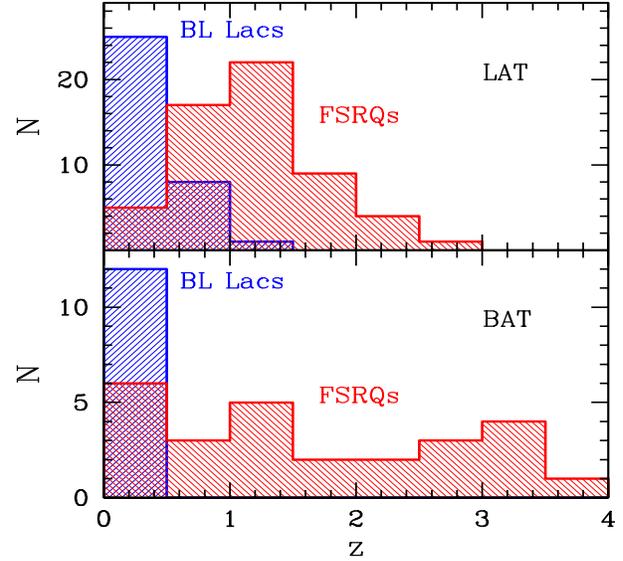,width=10cm,height=9cm}
\vskip -0.8 cm
\caption{Comparison of the redshift distributions of 
the bright {\it Fermi}/LAT and {\it Swift}/BAT
blazars, divided in BL Lacs and FSRQs.
}
\label{iz}
\end{figure}

Within the BAT sample, there are 10 blazars (all FSRQs) at redshift greater
than 2, and 5 at redshift between 3 and 4.
For comparison, the Large Area Telescope (LAT) onboard the {\it Fermi} satellite
detected about one hundred blazars (at high significance in the first 3 months), 
with a maximum redshift of 2.944 (Abdo et al. 2009).
Only 5 blazars have $z>2$, with only 1 with $z>2.5$.
Fig. \ref{iz} shows the distribution of redshifts for the {\it Fermi}/LAT and
the {\it Swift}/BAT blazars.
Comparing the two redshift distributions 
and including BL Lac objects, a K--S test gives a probability 
$P=0.079$ that they are drawn from the same distribution,
that decreases to $P=0.034$ when excluding BL Lacs.
The BAT FSRQs have a tail at high redshifts, not present in the case of LAT FSRQs.
This presence of this tail in the redshift distribution of BAT blazars
is the first indication that a survey in hard X--rays can be more fruitful 
in pinpointing the most powerful blazars lying at the highest redshifts.
Theoretically, this can be understood on the basis of the blazar sequence
(Fossati et al. 1998; Ghisellini et al. 1998, Donato et al. 2001; 
Ghisellini, Maraschi \& Tavecchio 2009).
According to this sequence the most powerful blazars, that are all FSRQs,
have the high energy peak of their spectral energy distribution (SED)
in the MeV range.
As a consequence, these objects are more luminous in hard X--rays than
what they are above 100 MeV, and thus become detectable in the BAT survey
even if they are undetected in the LAT one.
Indeed, {\it none} of the 10 BAT blazars at $z>2$ is present in the 3--months
LAT survey of bright blazars,
while 4 are present in the {\it Fermi}/LAT 11 months survey 
catalogue of sources with a $>4\sigma$ level 
detection\footnote{http://fermi.gsfc.nasa.gov/ssc/data/access/lat/1yr\_catalog/}
(PKS 0537--286, 0746+254, 0805+6144 and 0836+710),
and one additional source (PKS 2149--306) has a {\it Fermi}/LAT 
detection reported by B\"ock et al. (2010).
The {\it Fermi}/LAT fluxes of these sources are just above the limiting 
sensitivity for a 5$\sigma$ detection in 1 yr survey.

The aim of the present paper is to study the most powerful and distant blazars
present in the BAT survey in order to estimate the power carried by their
jets in the form of bulk motion of matter and fields.
Furthermore, we will estimate the mass of their central black hole and 
their accretion disk luminosities.
To this aim we will take advantage of the data of the other {\it Swift} 
instruments (the X--ray Telescope, XRT, and the UV and Optical Telescope, UVOT).
We will then compare the overall properties of these powerful BAT blazars
with the same properties of the bright $\gamma$--ray blazars detected by the LAT
and recently studied in Ghisellini et al. (2010, hereafter G10).

Finally, we will discuss the implications of our findings for the future 
missions, such as {\it NHXM} and {\it EXIST}, highlightening the possible 
discovery space that these missions can have concerning the search and the
study of the largest black hole masses of jetted sources at large redshift.

We use a flat cosmology with $h_0=\Omega_\Lambda=0.7$.
We adopt the convention $Q =10^x Q_x$ and use cgs units unless
specified otherwise.

\begin{table} 
\centering
\begin{tabular}{llllll}
\hline
\hline
Name        &Alias    &$z$    &$F_X$          &$\alpha_X$    &$\log L_X$ \\
\hline 
0014+813    &S5       &3.366  &1.88$\pm$0.21  &0.93$\pm$0.55 &48.03 \\
0222+185    &RBS315   &2.69   &1.42$\pm$0.22  &0.48$\pm$0.45 &47.63  \\
0537--286   &PKS      &3.10   &1.27$\pm$0.20  &0.56$\pm$0.30 &47.75  \\
074625+2549 &SDSS     &2.98   &1.49$\pm$0.25  &0.08$\pm$0.38 &47.50 \\
0805+6144   &GB6      &3.033  &0.96$\pm$0.19  &0.58$\pm$0.62 &47.62  \\
0836+710    &4C71.07  &2.17   &2.85$\pm$0.18  &0.47$\pm$0.14 &47.73  \\
1210+330    &B2       &2.50   &0.90$\pm$0.17  &0.40$\pm$0.30 &47.32  \\
2126--158   &PKS      &3.268  &1.55$\pm$0.27  &0.72$\pm$0.68 &47.99  \\
2149--306   &PKS      &2.35   &3.72$\pm$0.26  &0.52$\pm$0.21 &47.95 \\
225155+2217 &MG3      &3.668  &1.00$\pm$0.19  &0.51$\pm$0.33 &47.77  \\ 
\hline
\hline 
\end{tabular}
\vskip 0.4 true cm
\caption{
The 10 BAT blazars in A09 with $z>2$. 
The flux $F_X$ is in units of 10$^{-11}$ erg cm$^{-2}$ s$^{-1}$ and is 
calculated in the 15--55 keV energy range. 
The luminosity $L_X$ is calculated according to Eq. \ref{lx} and
is in units of erg s$^{-1}$.
}
\label{sample}
\end{table}

\section{The sample}

\begin{figure}
\vskip -0.6 cm \hskip -0.7 cm
\psfig{figure=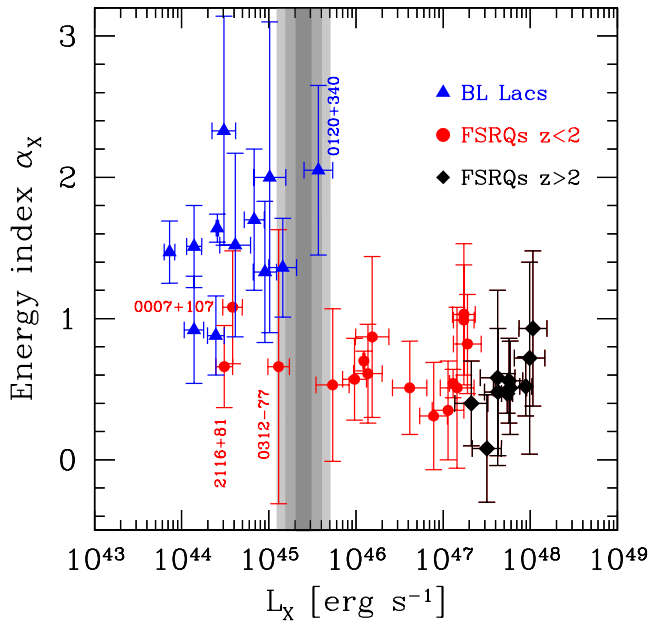,width=9.5cm}
\vskip -0.5cm
\caption{The X--ray energy spectral index $\alpha_X$ as a function
of the luminosity in the BAT energy range 15--55 keV.
In this plane BL Lacs (triangles) and FSRQs (circles and diamonds)
are clearly separated.
We also mark (black diamonds) the FSRQs with $z>2$.
The labelled sources are 
the BL Lac 1ES  0120+340 ($z=0.272$) and the three sources classified as FSRQs by A09
to the left of the divide.
Note that of these three objects, B0007+107 (= Mkn 1501) 
is a Seyfert 1 galaxy at $z=0.09$ with a superluminal jet 
(Brunthaler et al. 2000) and S5 2116+81 ($z=0.084$) is a radio--galaxy
(Stickel, K\"uhr \& Fried 1993).
The grey stripes indicate the X--ray luminosity dividing
BL Lacs from FSRQs. 
}
\label{algamma}
\end{figure}

In A09 all BAT blazars were fitted with a simple power law  
in the 15--55 keV energy range. 
The resulting energy spectral indices $\alpha_X$ are shown in 
Fig. \ref{algamma} as a function of the rest frame [15--55 keV] 
luminosity, calculated according to:
\begin{equation}
L_X\, =\, 4\pi d_{\rm L}^2 {F_X \over (1+z)^{1-\alpha_X} } 
\label{lx}
\end{equation}
where $F_X$ is the observed X--ray flux in the 15--55 keV energy range
as listed in A09.

Blazars with $z>2$ are marked with black diamonds:
they are the most luminous, with $L_X>2\times 10^{47}$ erg s$^{-1}$
in the (rest frame) 15--55 keV band. 
Our cut in redshift therefore corresponds also to a cut in luminosity,
as expected for a flux limited sample.

Table \ref{sample} reports the redshift, the BAT X--ray flux in 
the 15--55 keV range, 
the energy spectral index in this energy range and the corresponding
K--corrected X--ray luminosity. 
Is should be noted that these 
10 blazars {\it are not} the only $z>2$ blazars detected by BAT.
Indeed, there are 4 additional blazars
in the $|b|<15^\circ$ region
(0212+735, with $z=2.367$; OA 198 at $z=2.365$;
SWIFT J1656.3--3302 at $z=2.4$ and PKS 1830--211 at $z=2.507$),
which are present in the catalogues 
of Cusumano et al. (2009) and of Tueller et al. (2009)
We chose  the A09 catalogue mainly 
because of their derivation of the blazar luminosity function
performed with the data in that paper.

Several of the sources listed in Tab. \ref{sample} have already been discussed
and modelled in the literature: S5 0014+813 was discussed by Ghisellini et al. (2009) 
because of its apparently huge black hole mass (40 billion solar masses, but
see Ghisellini et al. 2009 for the possibility that this is over--estimated 
because of the assumption of an isotropically emitting accretion disk,
and see below for further discussion);
RBS 315 was discussed in Tavecchio et al. (2007, presenting also Suzaku data),
PKS 0537--286 is discussed in Sambruna et al. (2007) and
in Bottacini et al. (2010, presenting also {\it INTEGRAL} data);
J074625+2549 is discussed in Sambruna et al. (2006) and Watanabe et al. (2009, 
presenting also
{\it Suzaku} data);
4C 71.07 (=0836+710) is discussed in 
Foschini et al. (2006) and
Sambruna et al. (2007);
PKS 2149--306 is discussed in Sambruna et al. (2007) and in Bianchin et al. (2009, 
presenting also XMM--{\it Newton} and {\it INTEGRAL} data);
J225155+2217 is discussed in Bassani et al. (2007, presenting also {\it INTEGRAL} data),
and in Maraschi et al. (2008).

To the best of our knowledge, the overall SED of the remaining 3 blazars:
GB6 J0805+6144, B2 1210+330 and PKS 2126--158
have not yet been discussed and modelled.

\subsection{The BAT blazars' divide}

Fig. \ref{algamma} shows that BL Lacs and FSRQs are clearly separated
in the $\alpha_X$--$L_X$ plane: BL Lacs have steeper slopes and lower luminosities
than FSRQs. 
This can be readily explained recalling that, in BL Lacs,
the X--ray flux often belongs to the tail of the synchrotron hump,
while in more powerful FSRQs the X--ray flux always belongs to the 
very hard portion of the high energy hump, believed to be produced by
Inverse Compton process.
The separation occurs at $\alpha_X\sim 1$ and at $L_X\sim 10^{45}$ erg s$^{-1}$.
This ``divide" is related to a similar divide present in the
$\alpha_\gamma-L_\gamma$ plane (Ghisellini, Maraschi \& Tavecchio 2009), 
using the 3--months data of the {\rm Fermi}/LAT survey.
However, there is one important difference:
in the $\alpha_\gamma-L_\gamma$ plane BL Lacs are the {\it hard} sources,
while FSRQs are the softer ones. 
In the $\gamma$--ray energy range the flux
originates from the same (Inverse Compton) component, and
low luminosity BL Lacs peaks at energies close to the 
end of the LAT energy range, or even at higher energies,
and therefore have $\gamma$--ray slopes flatter than FSRQs,
which instead peaks at energies smaller than 100 MeV.
Fig. \ref{algamma} is therefore another manifestation of the
blazar sequence, because it agrees with the ideas that low power 
blazars have both the synchrotron and the Compton peaks at larger 
frequencies than FSRQs. 
In this sense we have ``blue" BL Lacs and ``red" FSRQs.
We propose here that
the existence of a dividing X--ray luminosity is the same as discussed
in Ghisellini, Maraschi \& Tavecchio (2009), namely a change in the accretion
regime of the underlying accretion disk.
This interpretation assumes that there is a connection between the non--thermal
beamed luminosity and the thermal luminosity produced by accretion, and that
the range of black hole masses of the blazars illustrated in Fig. \ref{algamma}
is relatively small.
Within this context, the most luminous objects are the ones accreting 
close to the Eddington rate.
The dividing X--ray luminosity is produced by objects whose accretion disk
is accreting at a rate a factor $\sim 300$ lower. 
If these sources have the same mass as the most powerful blazars, then they
accrete at the 0.3\% of the Eddington rate, and at these rates the accretion
could change regime, becoming radiatively inefficient (see the discussion
in Ghisellini, Maraschi \& Tavecchio 2009).
In turn, the rather abrupt decrease of ionising photons occurring at and 
below the critical accretion rate implies an equally abrupt decrease of the
broad line luminosity, making the blazar appear as a line--less BL Lac.

\section{Swift data and analysis}

We collected the public {\it Swift} data of the 10 sources listed in 
Tab. \ref{sample}. 
For all but one source, B2 1210+330, we could find at least 
one observation in the Swift archive. 
We obtained a {\it Swift} ToO for this source: 
the observations were performed on the 26th and
27th of October 2009. 

\subsection{XRT data}

The XRT data were processed with the standard procedures 
({\texttt{XRTPIPELINE v.0.12.2}). 
We considered photon counting (PC) mode data with the standard 0--12 grade selection. 
Source events were extracted in a circular region of  aperture $\sim 47''$, 
the background was estimated in a same sized circular region far from the source. 
Response matrices were created through the \texttt{xrtmkarf} task. 
The channels with energies below 0.3 keV and above 10 keV were excluded 
from the fit and the spectra were rebinned in energy so to have at 
least 30 counts per bin.
For spectra with very few counts, the Cash statistics was applied.
When a sufficiently long exposure is available, we present the 
data corresponding to that observations.
In a few cases we summed the data of different observations, 
to have a more accurate fit, after having checked that the 
fits of the single observations gave consistent results.

Each spectrum was analysed through XSPEC
with an absorbed power--law using the Galactic column density 
($N^{\rm Gal}_{\rm H}$ from Kalberla et al. 2005). 
The computed errors represent the 90\% confidence interval on the spectral parameters.
In Table \ref{xrt} we list the results.

\subsection{UVOT data}

UVOT (Roming et al. 2005) source counts were extracted from 
a circular region $5''-$sized centred on the source position, 
while the background was extracted from 
a larger circular nearby source--free region.
Data were integrated with the \texttt{uvotimsum} task and then 
analysed by using the  \texttt{uvotsource} task.

The observed magnitudes have been dereddened according to the formulae 
by Cardelli et al. (1989) and converted into fluxes by using standard 
formulae and zero points from Poole et al. (2008).
No further absorption was applied. 
Tab. \ref{uvot} lists the result.

\begin{table*} 
\centering
\begin{tabular}{llllllllll}
\hline
\hline
\hline   
source	&OBS date   &$t_{\rm exp}$ &$N^{\rm Gal}_{\rm H}$  &$\Gamma_1$ &$\Gamma_2$ &$E_{\rm break}$ &$F_{\rm 0.3-10}^{\rm unabs}$ &$\chi^2$ or C$_{\rm stat}$ &dof    \\
        &dd/mm/yyyy &$s$           &$10^{20}$ cm$^{-2}$    &           &           &keV             &10$^{-12}$ cgs & &  \\   
\hline
0014+81    &sum$^{(a)}$        &8311  &13.6 &1.32$\pm$0.1  &              &   &5.4          &22.6        &19    \\ 
0222+185   &28/07/2006 &4150  &9.3  &1.20$\pm$0.08 &              &   &14$\pm$8     &344         &378  \\
0537--286  &08/12/2005 &14737 &2.2  &1.20$\pm$0.08 &              &   &4$\pm$0.1    &333         &407  \\
074625+2549&05/11/2005 &24999 &4.6  &1.22$\pm$0.06 &              &   &4.6$\pm$0.2  &467         &543 \\
0805+614   &sum$^{(b)}$        &8121  &4.7  &1.25$\pm$0.13 &              &   &3.14         &10.8        &10 \\ 
0836+710   &13/04/2007 &7367  &2.9  &1.42$\pm$0.05 &              &   &17.5$\pm$0.5 &526         &532  \\
1210+330   &26/10/2009 & 3782 &1.16 & 1.9$\pm0.37$     &              &   &0.57$\pm$0.15 & 31             & 41    \\
2126--158  &sum$^{(c)}$	     &34782 &5.0  &0.6$\pm$0.3   &1.5$\pm$0.05  &0.95$\pm$0.2&10.3 &204        &161\\ 
2149--306  &10/12/2005 &3336  &1.6  &1.45$\pm$0.07 &              &                  &16.2$\pm$8 &367   &394  \\
225155+2217&22/05/2007 &12396 &5.0  &1.28$\pm$0.08 &              &	               &4.66 &19   &25\\
\hline
\hline
\end{tabular}
\vskip 0.4 true cm
\caption{Results of the X--ray analysis. (a) sum of the observations 11/01/2007, 
12/01/2007 and 14/01/2007; (b) sum of the observations 23/01/2009, 17/10/2006 and 21/01/2009; 
(c) sum of the observations 07/04/2007, 21/11/2008 and 06/04/2007.
}
\label{xrt}
\end{table*}

\begin{table*}
\centering
\begin{tabular}{llllllll}
\hline
\hline
Source       &$A_V$ &$v$              &$b$             &$u$            &$uvw1$         &$uvm2$         &$uvw2$\\
\hline
0014+81      &0.70   &$16.43\pm 0.05$ &$17.57\pm 0.06$ &$18.2\pm 0.1$  &$>19.4$        &$>19.6$        &$>20.0$\\
0222+185     &0.916  &$18.62\pm 0.25$ &$19.28\pm 0.13$ &$19.35\pm0.18$ &$>20.65$       & ...           &$>20.85$ \\
0537--286    &0.125  &$19.39\pm 0.1$  &$19.94\pm0.07$  &$21.05\pm0.19$ &$>22.25$       &$>22.31$       &$>22.82$ \\
074625+2549  &0.12   &$19.42\pm0.14$  &$19.84\pm0.14$  &$20.09\pm0.17$ &$>21.6$        &$>21.89$       &$>22.37$ \\
0805+6144    &0.189  &$>19.3$         &$>20.4$         &$>20.1$        &$>20.5$        &$>20.3$        &$>20.4$ \\
0836+710     &0.101  &$17.17\pm0.05$  &$17.29\pm0.03$  &$16.46\pm0.02$ &$17.34\pm0.03$ &$17.88\pm0.05$ &$17.21\pm0.04$ \\
1210+330     &0.041  &$19.01\pm0.25$   &$18.97\pm0.11$    &$18.78\pm0.12$   &$19.38\pm0.14$   &$19.00\pm0.11$   &$20.13 \pm0.17$ \\
2126--158    &0.264  &$17.03\pm 0.06$ &$18.20\pm 0.06$ &$19.5\pm 0.2$  &$>19.8$        &$>20.3$        &$>20.6$        \\
2149--306    &0.083  &$17.66\pm$0.08  &$17.88\pm0.06$  &$17.33\pm0.05$ &$18.31\pm0.08$ &$20.44\pm0.24$ &$20.21\pm0.13$ \\
225155+2217  &0.272  & ...            & ...            &$>20.7$        &$>22.0$        &$>22.2$        &$>21.2$          \\
\hline
\hline
\end{tabular}
\caption{Summary of \emph{Swift}/UVOT observed magnitudes. 
Lower limits are at $3\sigma$ level. 
Values of $A_V$ from Schlegel et al. (1998).
}
\label{uvot}
\end{table*}

\section{Lyman--$\alpha$ and Lyman--edge absorption}

Being at redshifts between 2 and $\sim$3.7, the optical--UV flux of
the blazars in our sample could be affected by absorption
of neutral hydrogen in intervening Lyman--$\alpha$
absorption systems. 
Single ionised helium is not an issue at these redshifts 
for UVOT wavelengths. 
Not knowing the real attenuation along individual line of sights, 
we estimate the {\it probable} attenuation using the effective opacity 
$\tau_{\rm eff}\equiv -\ln<{\rm e}^{-\tau}>$, where the average is taken over all 
possible line of sights. 
We compute the mean attenuation for six wavelengths 
approximately centred in the six UVOT filters.
Fig. \ref{tau} shows the corresponding $\tau_{\rm eff}$
as a function of redshift. 
We adopt the column density distribution described in Haardt \& Madau 
(in preparation), 
which is based on the mean free path measurements of 
Prochaska et al. (2009). 
Full details os our calculation will be described in Haardt et al. 
(in preparation), together with a more refined treatment of the mean 
attenuation and its variance around the mean.
Such procedure is very crude, as the attenuation variance along an 
individual line of sight is large for a fixed observed wavelength. 
We must note however that: 
i) the variance of the attenuation is largely reduced when the actual filter width is 
taken into account (Madau 1995), and 
ii) the variance of the attenuation is mainly driven by the few Lyman 
limit systems present along an 
individual line of sight. 
%
\begin{figure}
\vskip -0.6 cm \hskip -0.8 cm
\psfig{figure=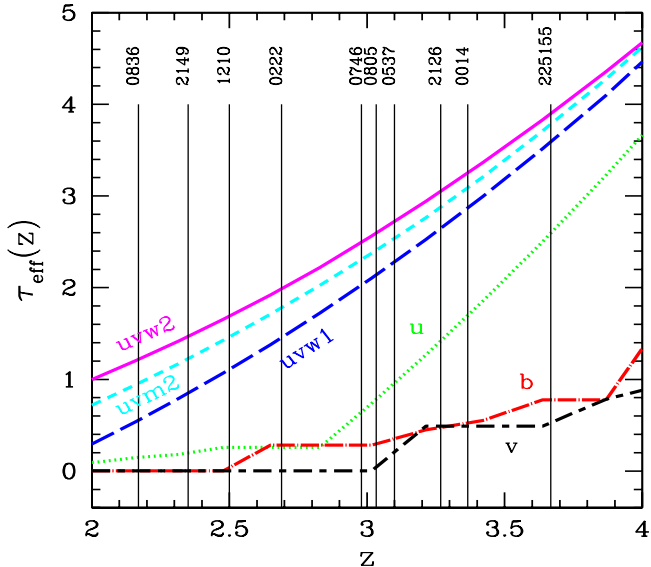,width=10cm,height=9cm}
\vskip -0.8 cm
\caption{ 
The optical depth due to neutral hydrogen and helium
as a function of redshifts. Vertical lines (and labels)
indicate the redshifts of the blazar in our sample.
The different curves are for the 6 filters of UVOT,
as labelled.
}
\label{tau}
\end{figure}
%
Our absorption model results in a mean number of thick systems which is
$<1$ for $z\lsim 4$, so we do note expect excessive off--set of the 
attenuation along individual line of sight with respect to the mean value.

When presenting the SED of our sources, we will show 
both the fluxes and upper limits de--reddened for the extinction 
due to our Galaxy and the fluxes (and upper limits) obtained
by de--absorbing them with the $\tau_{\rm eff}$ shown in Fig. \ref{tau}.

\begin{figure}
\vskip -0.6 cm \hskip -0.4 cm
\psfig{figure=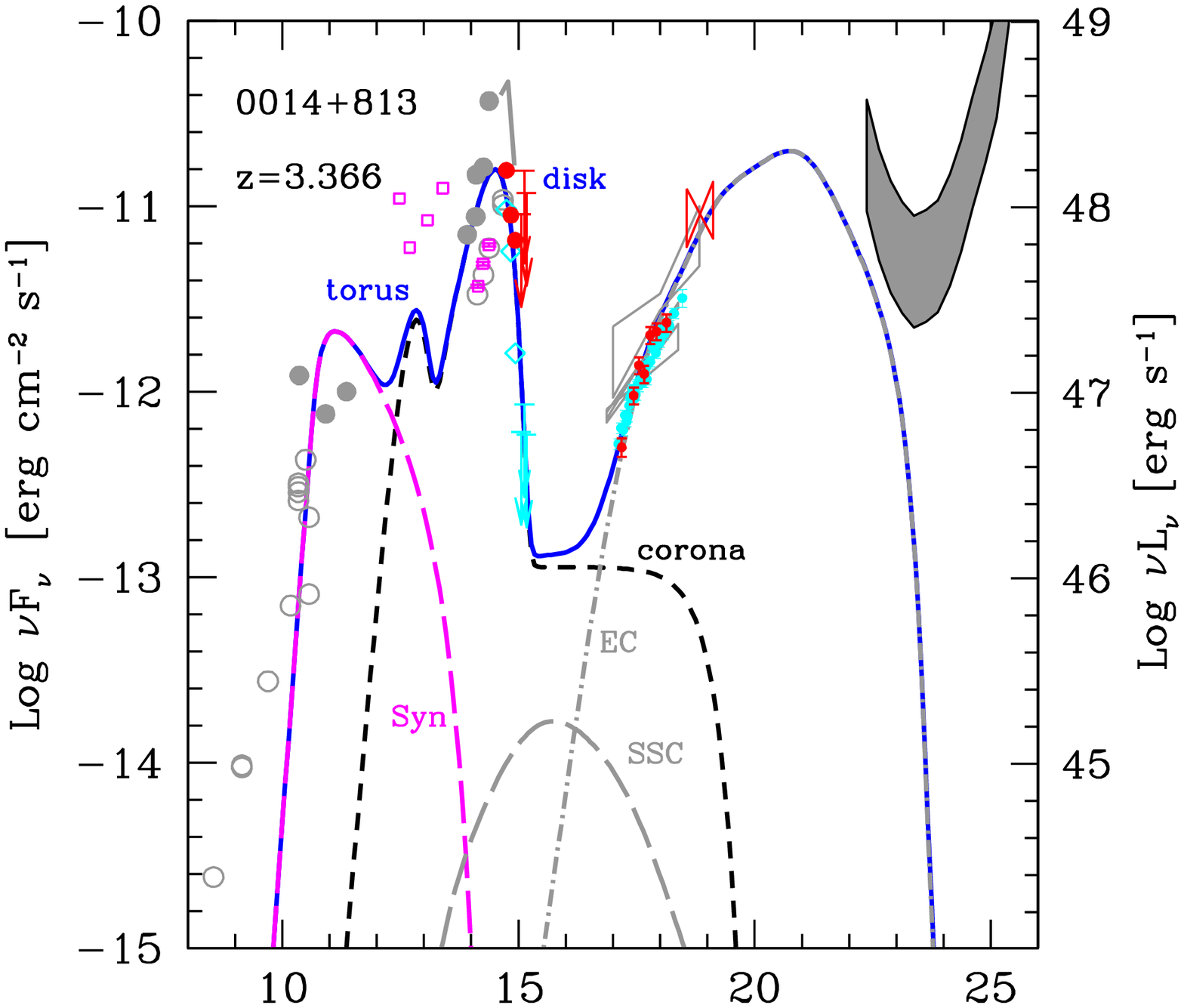,width=9cm,height=6.9cm}
\vskip -1.3 cm \hskip -0.4 cm
\psfig{figure=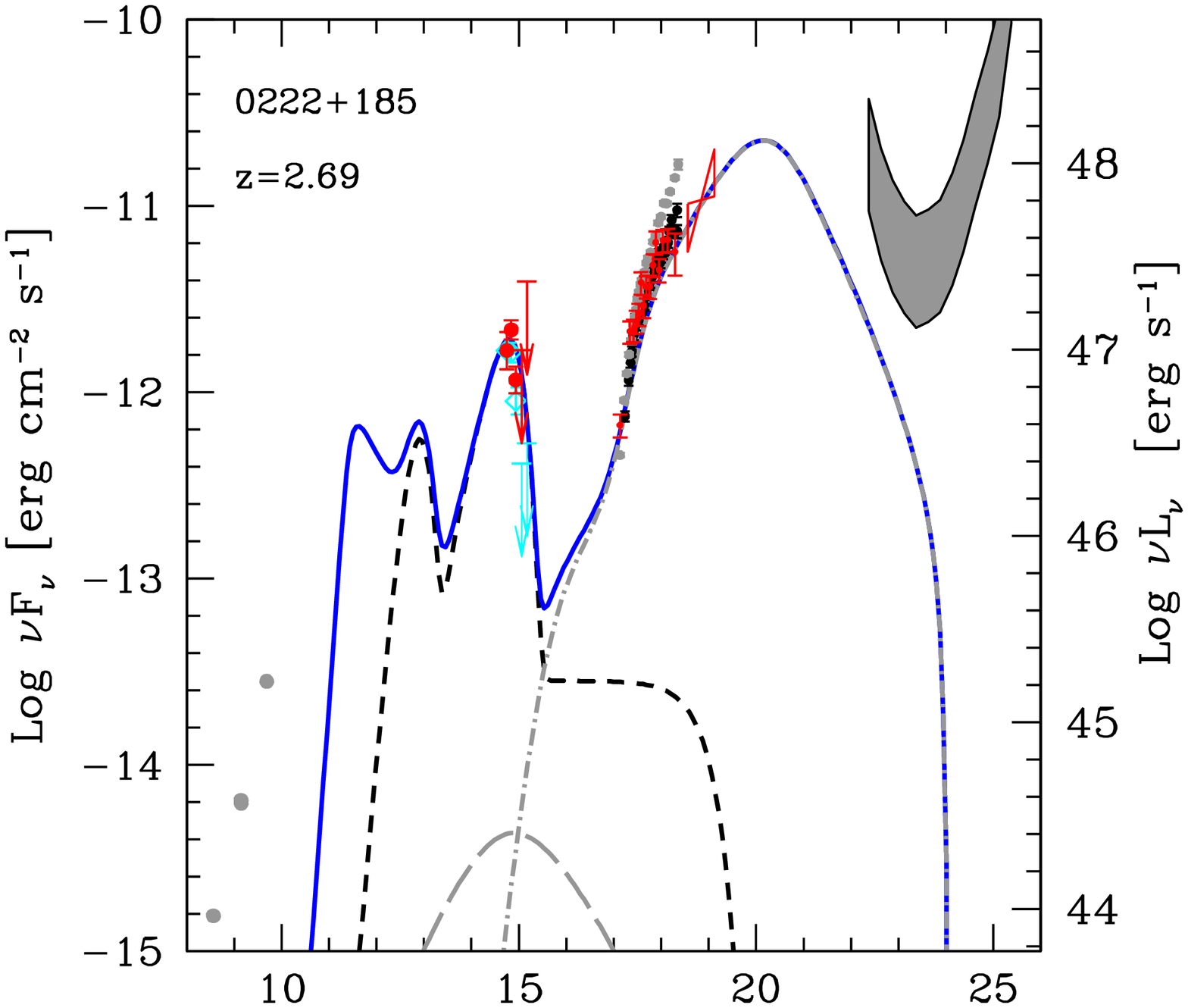,width=9cm,height=6.9cm}
\vskip -1.3 cm \hskip -0.4 cm
\psfig{figure=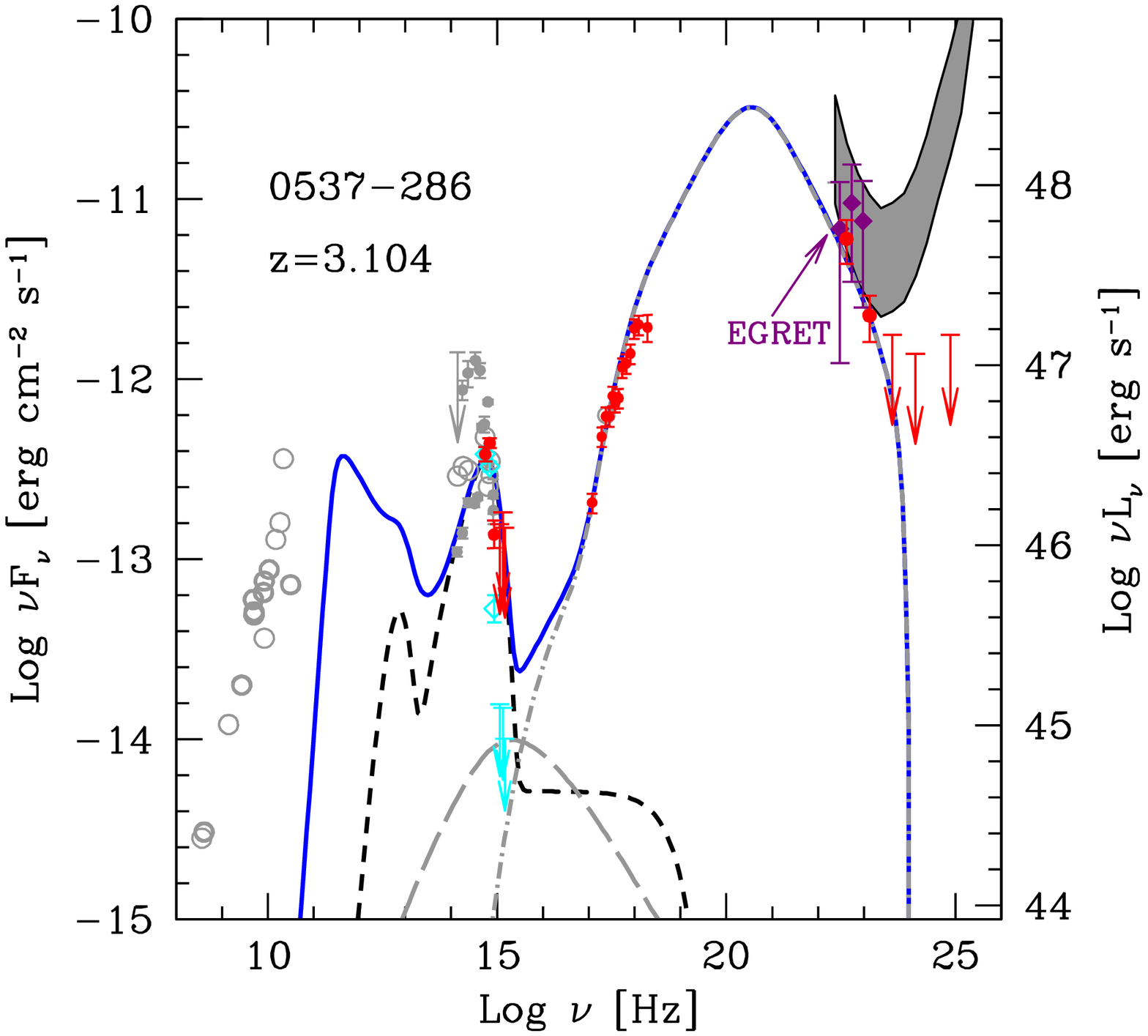,width=9cm,height=6.9cm}
\vskip -0.8 cm
\caption{
SEDs of S5 0014+813, 0222+185 (=RBS 315) and PKS 0537--286
together with the fitting models,
with parameters listed in Tab. \ref{para}.
De--absorbed UVOT, XRT and BAT data are indicated
by darker symbols (red in the electronic version),
while archival data (from NED) are in light grey.
Diamonds (and lower arrows, cyan in the electronic version) 
indicate UVOT data not de--absorbed by
intervening Lyman--$\alpha$ clouds.
The magenta square symbols for S5 0014+813
are IRAS and 2MASS data points.
The short--dashed line is the emission from the IR torus, 
the accretion disk and its X--ray corona. 
The long dashed and the dot--dashed grey lines are the synchrotron self--Compton 
(SSC) and the External Compton (EC) components, respectively. 
The thick solid (blue) line is the sum of all components.
}
\label{f1}
\end{figure}

\begin{figure}
\vskip -0.6cm \hskip -0.4 cm
\psfig{figure=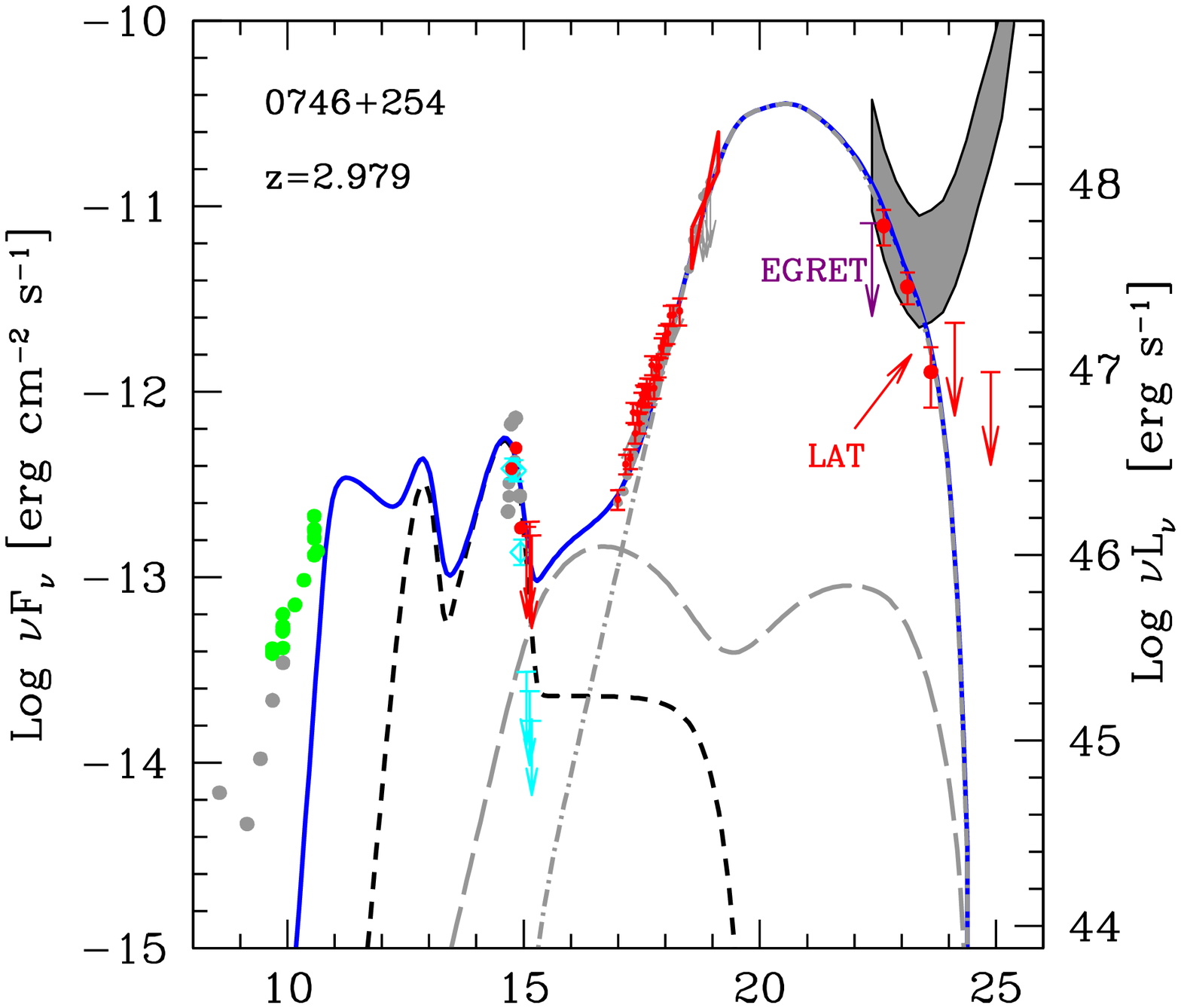,width=9cm,height=6.9cm}
\vskip -1.3 cm \hskip -0.4 cm
\psfig{figure=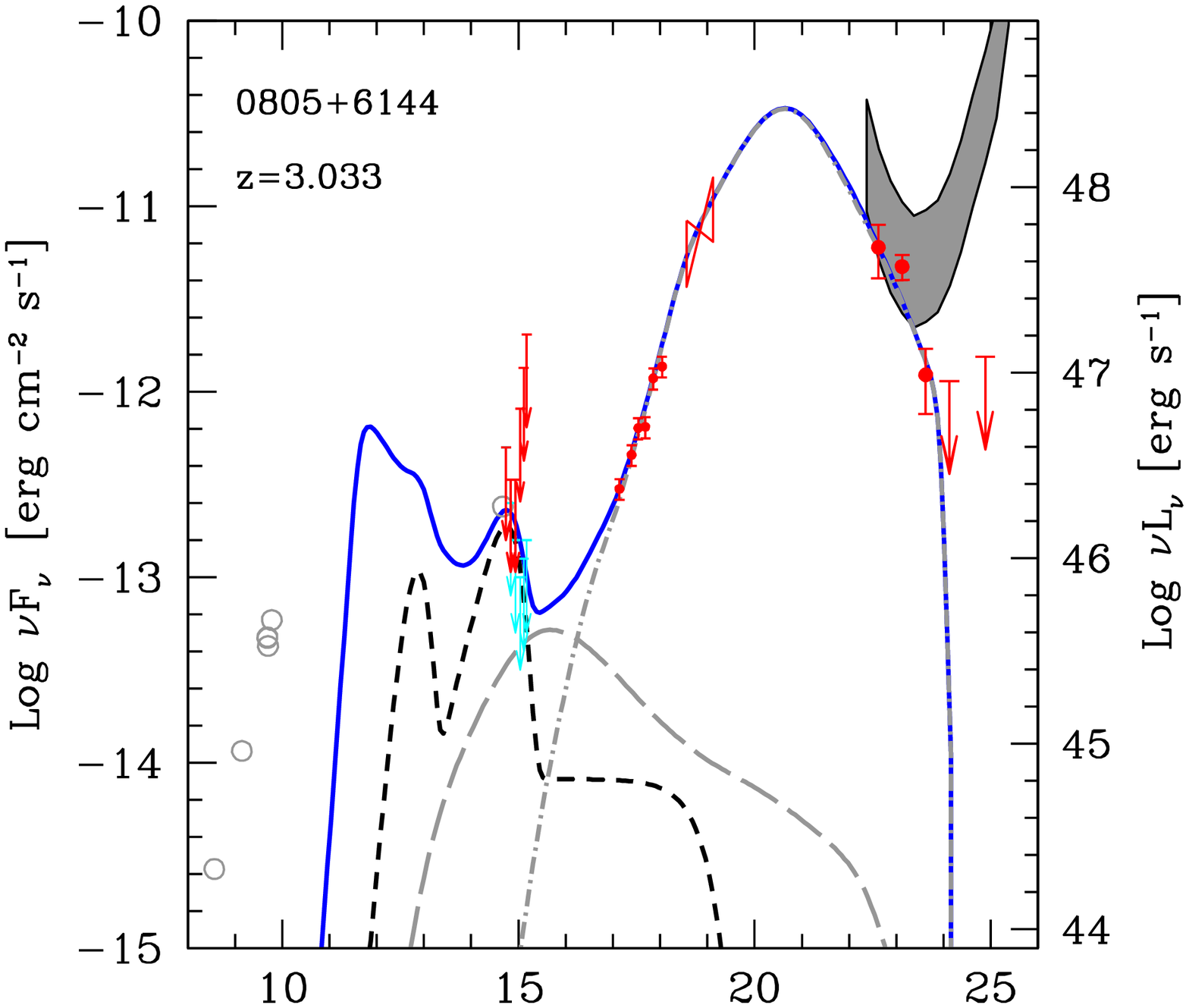,width=9cm,height=6.9cm}
\vskip -1.3 cm \hskip -0.4 cm
\psfig{figure=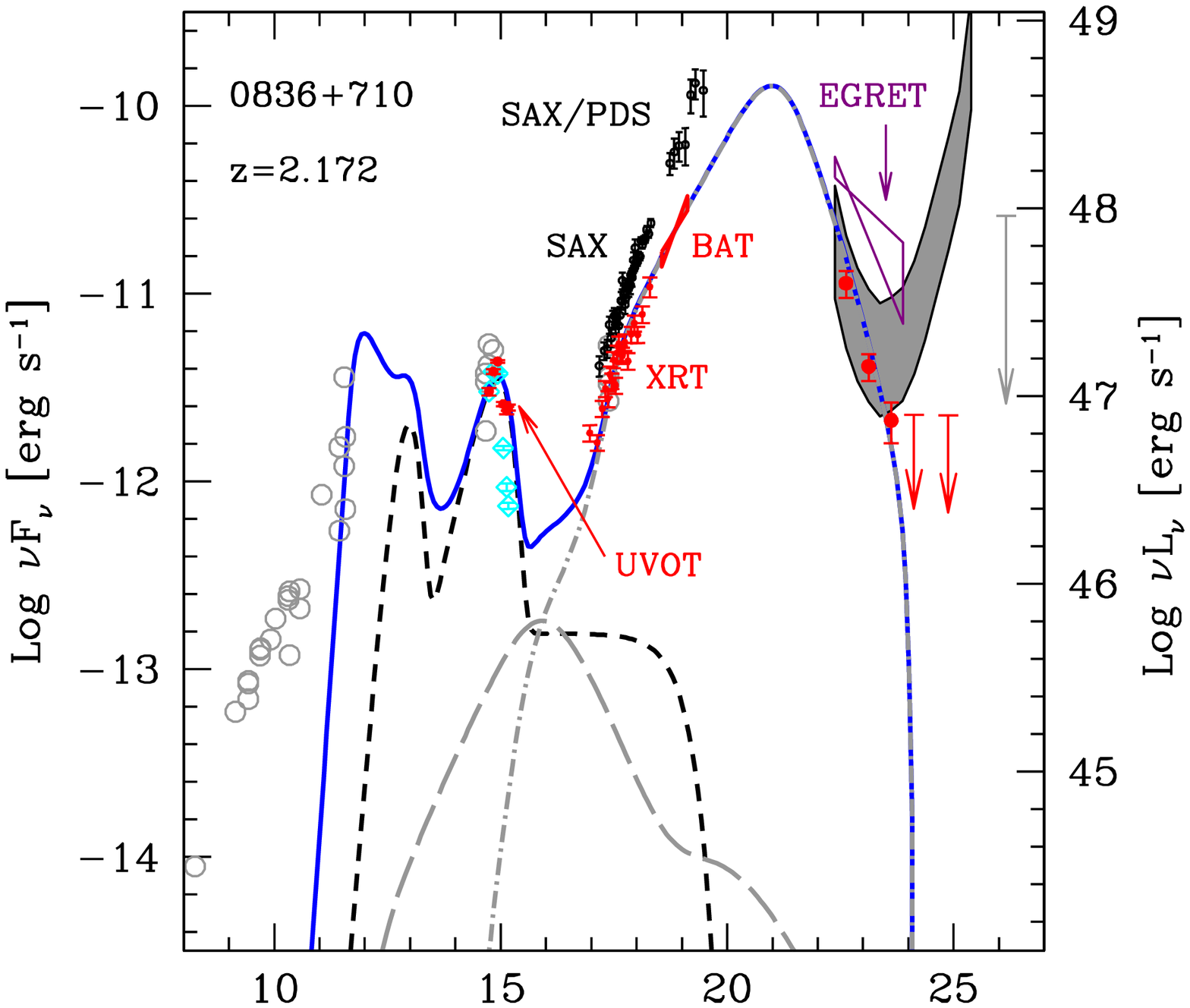,width=9cm,height=6.9cm}
\vskip -1.3 cm \hskip -0.4 cm
\psfig{figure=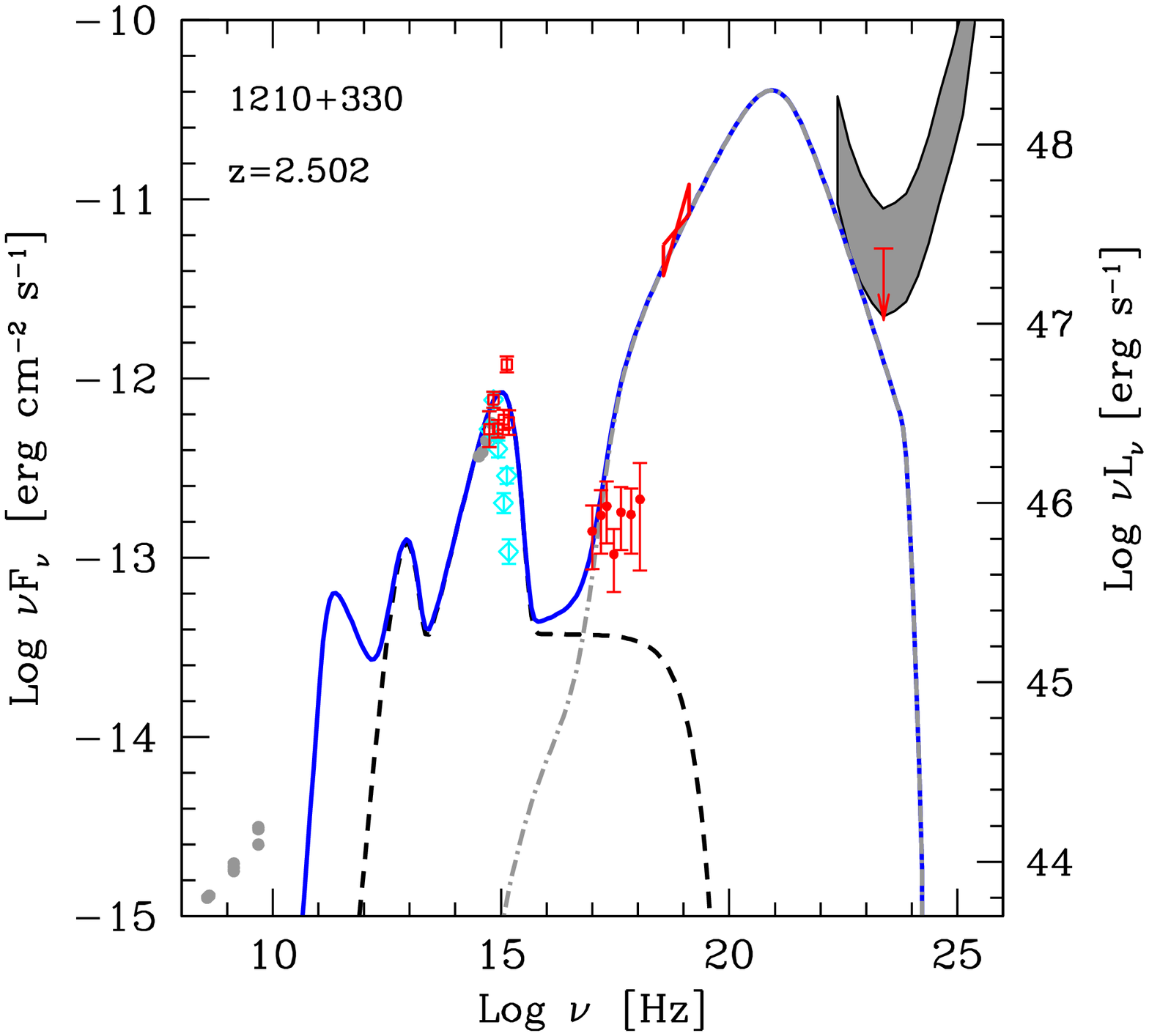,width=9cm,height=6.9cm}
\vskip -0.8 cm
\caption{
SED of SDSS 074625+2549, PKS 0805-6144, PKS 0836+710 and B2 1210+330. 
Symbols and lines as in Fig. \ref{f1}.
}
\label{f2}
\end{figure}

\section{The spectral energy distribution}

Although the SED of several blazars of our sample have already been 
presented in the literature, we show them all in Fig. \ref{f1}--\ref{f3},
adding to the archival data the BAT data from the A09 catalogue.
The SEDs show the following general properties:
\begin{itemize}
\item 
We show as a grey stripe the detection limits of the LAT instrument:
the lower bound corresponds to the pre--flight 
differential sensitivity limit 
for one year survey at the 5 $\sigma$ level\footnote{see:
http://fermi.gsfc.nasa.gov/ssc/data/analysis/documentation\\/Cicerone/Cicerone\_LAT\_IRFs/LAT\_sensitivity.html
}, 
while the upper bound is simply a factor 4 higher.
The latter should mimic the approximate sensitivity limit
for the three months LAT survey at 10$\sigma$ confidence level.
These curves depend on the assumed spectral index of the source,
and have been obtained assuming $\alpha_\gamma=1$, approximately
valid for the expected spectrum of FSRQs.
All sources in the present sample have not been detected in
the first 3--months LAT survey. 
Their $\gamma$--ray flux should
then lie below the upper bound of the grey stripe.
This, together with the BAT data, constrains   
the high energy peak of the SED to occur in the MeV band. 
Two blazars, 0537--286 and 0836+710, were detected by EGRET
(Nandikotkur et al. 2007; Thompson et al. 1993)
with a flux level reported in Fig. \ref{f1} and Fig. \ref{f2}.
This indicates a large variability amplitude.

\item The high energy peak is dominating the electromagnetic output 
with luminosities exceeding $10^{48}$ erg s$^{-1}$ in all cases
but S5 0014+813 (in which it is the accretion disk luminosity 
that dominates the power output).

\item The 0.3--10 keV {\it Swift}/XRT data indicate a very hard X--ray spectrum,
harder than the BAT one, but, in general, the extrapolation from the XRT agrees
with the BAT flux.
The BAT spectrum is a three--years average, while the XRT
spectrum is more often a one--epoch spectrum, and rarely an average of
the available observations.

\item In the optical--UV band the spectrum is almost always steep
(exceptions are 1210+330 and 225155+2217), and sometimes shows a peak.
This component can be interpreted as the emission from the accretion disk
(see G10; Ghisellini et al. 2009, and \S 4.2).

\item When there are multiple X--ray observations,
these show variability, with an apparent tendency for a ``harder when brighter"
behaviour of the X--ray flux 
(see 0222+185; 0836+710; 2126--158; 2149--306 and 225155+2217).
The variability amplitude thus is greater at higher energies,
being very modest below a few keV.

\end{itemize}
%

\begin{figure}
\vskip -0.6 cm \hskip -0.4 cm
\psfig{figure=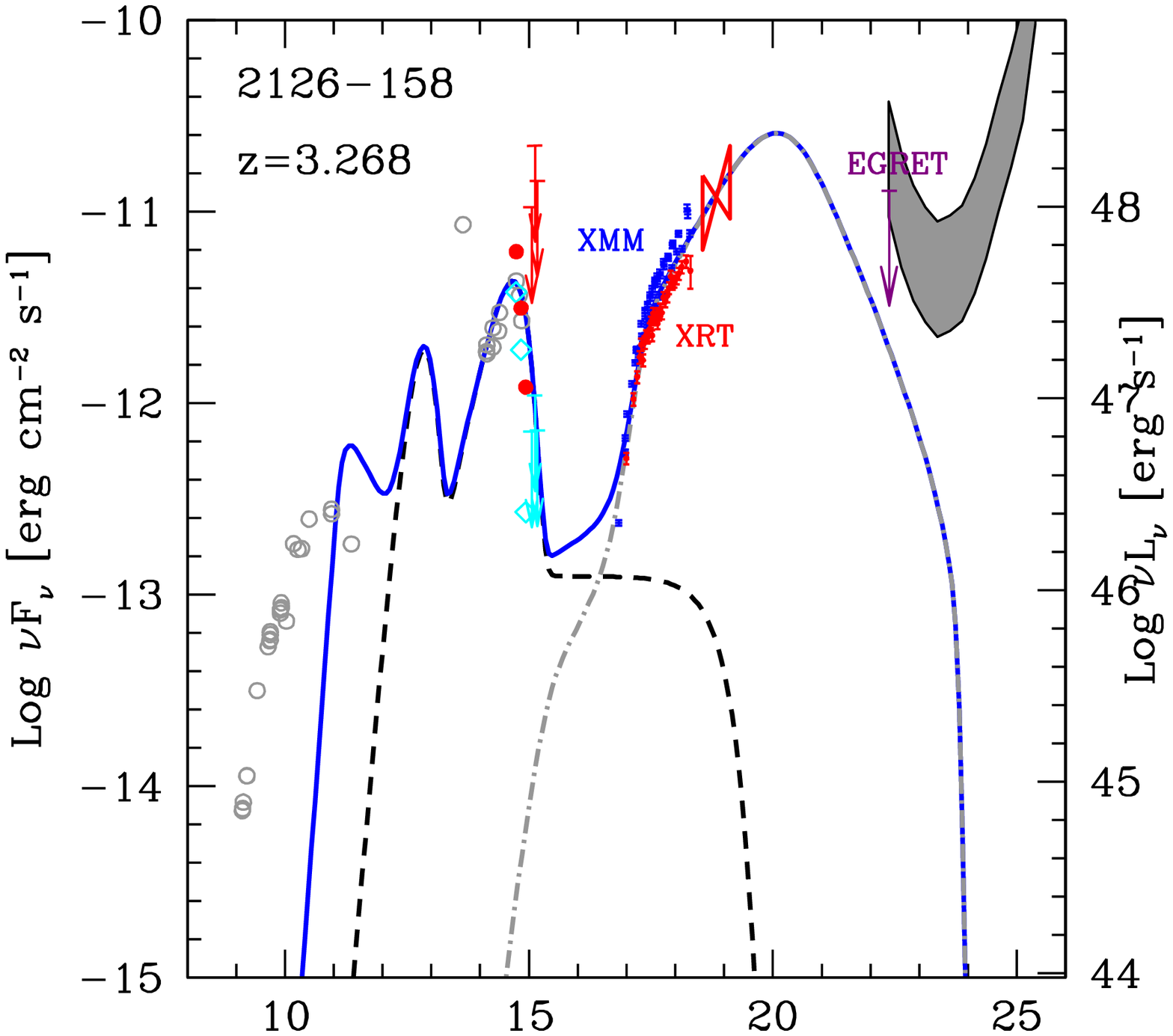,width=9cm,height=7cm}
\vskip -1.3 cm \hskip -0.4 cm
\psfig{figure=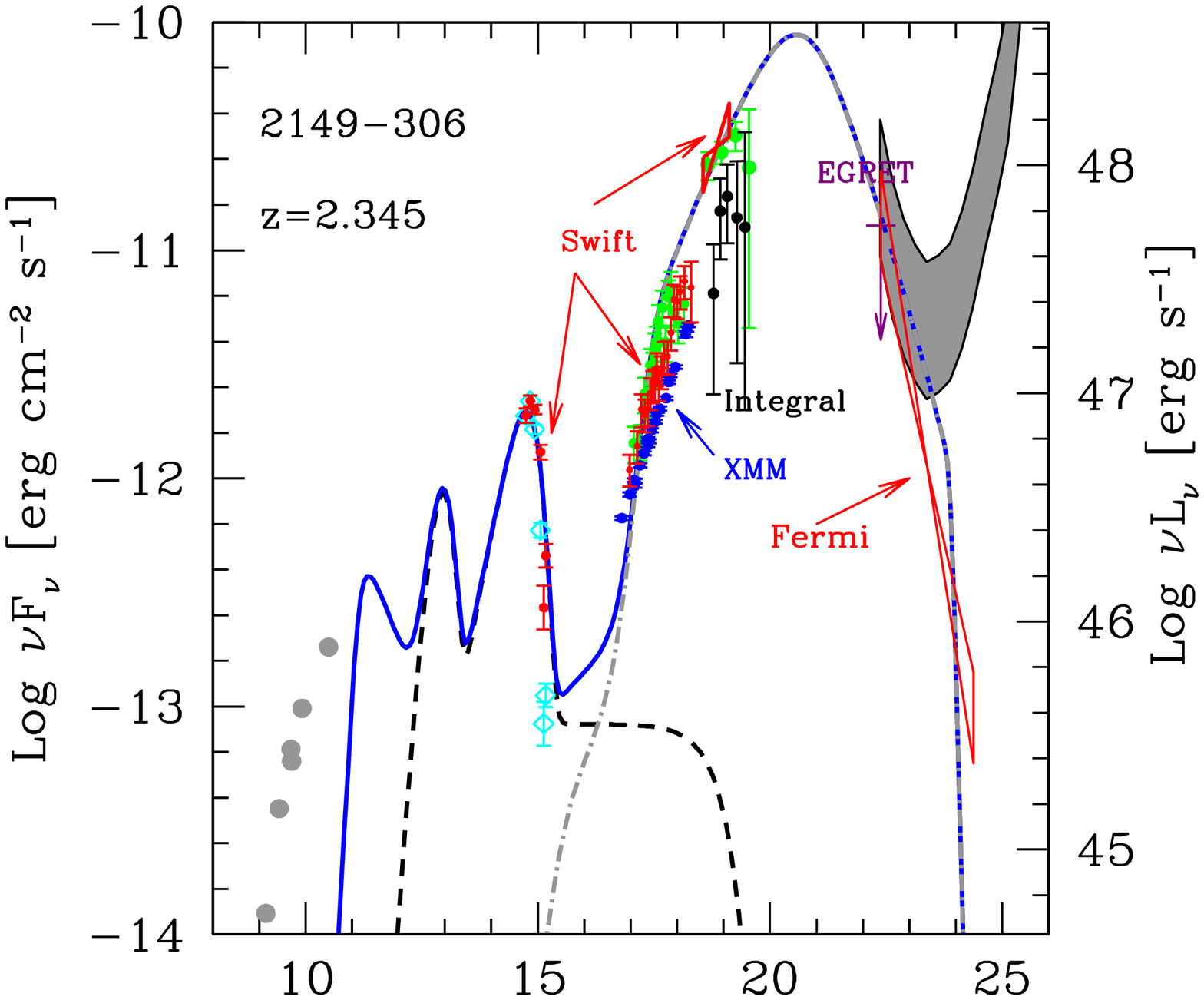,width=9cm,height=7cm}
\vskip -1.3 cm \hskip -0.4 cm
\psfig{figure=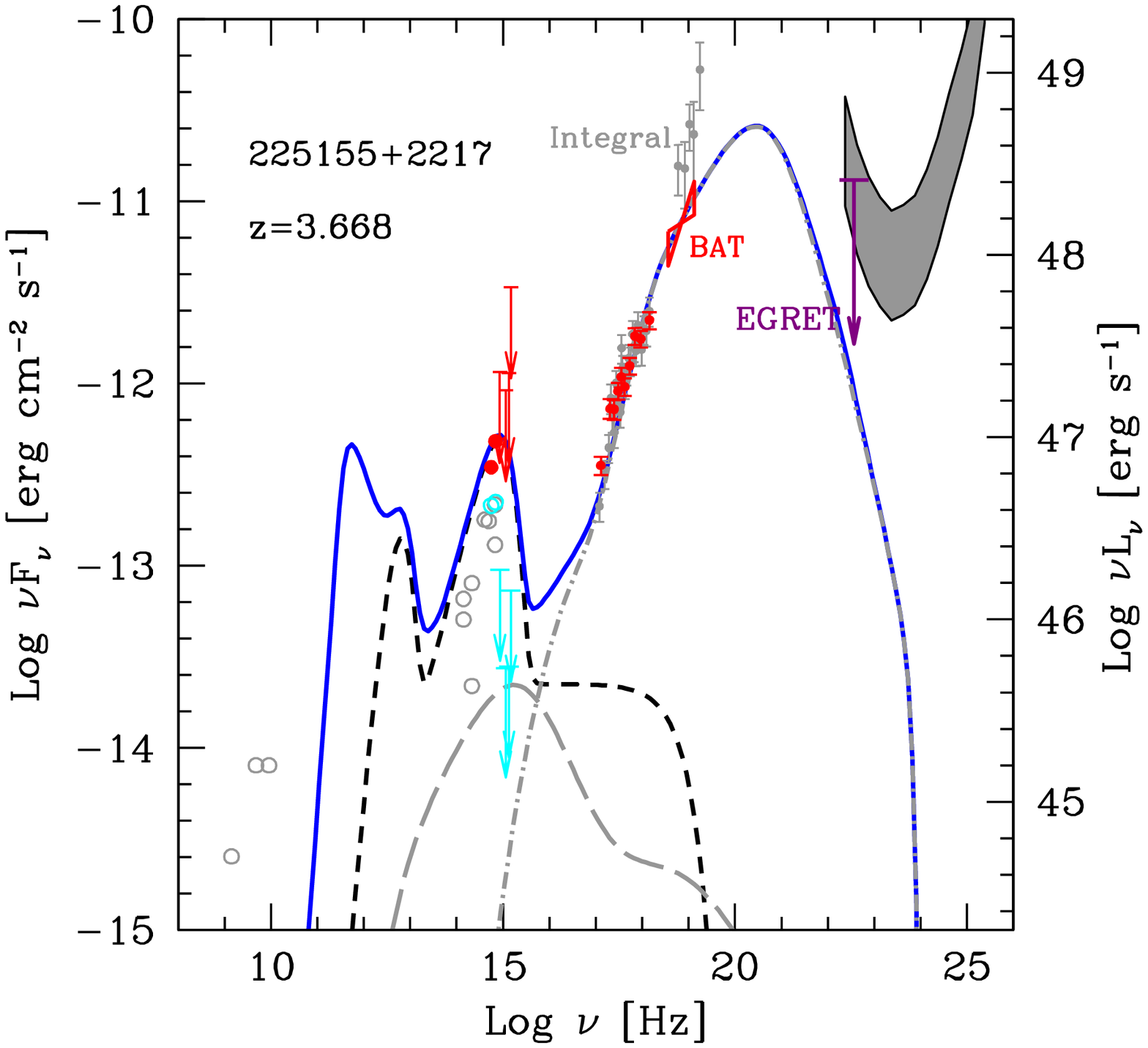,width=9cm,height=7cm}
\vskip -0.8 cm
\caption{
SED of PKS 2126--158,
PKS 2149--306 and  MG3 J225155+2217.
Symbols and lines as in Fig. \ref{f1}.
}
\label{f3}
\end{figure}

\subsection{The model}

To model the SED we have used a relatively simple leptonic,
one--zone synchrotron and inverse Compton model.
This model, fully discussed in Ghisellini \& Tavecchio (2009),
has the following main characteristics.

We assume that in a spherical region of radius $R$, located at a distance
$R_{\rm diss}$ from the central black hole, relativistic electrons are injected at
a rate $Q(\gamma)$ [cm$^{-3}$ s$^{-1}$] for a finite time equal to the 
light crossing time $R/c$. 
For the shape of $Q(\gamma)$ we adopt a smoothly broken power law,
with a break at $\gamma_{\rm b}$:
\begin{equation}
Q(\gamma)  \, = \, Q_0\, { (\gamma/\gamma_{\rm b})^{-s_1} \over 1+
(\gamma/\gamma_{\rm b})^{-s_1+s_2} }
\label{qgamma}
\end{equation}

The emitting region is moving with a  velocity $\beta c$
corresponding to a bulk Lorentz factor $\Gamma$.
We observe the source at the viewing angle $\theta_{\rm v}$ and the Doppler
factor is $\delta = 1/[\Gamma(1-\beta\cos\theta_{\rm v})]$.
The magnetic field $B$ is tangled and uniform throughout the emitting region.
We take into account several sources of radiation externally to the jet:
i) the broad line photons, assumed to re--emit 10\% of the accretion luminosity
from a shell--like distribution of clouds located at a distance 
$R_{\rm BLR}=10^{17}L_{\rm d, 45}^{1/2}$ cm;
ii) the IR emission from a dusty torus, located at a distance
$R_{\rm IR}=2.5\times 10^{18}L_{\rm d, 45}^{1/2}$ cm;
iii) the direct emission from the accretion disk, including its X--ray corona;
iv) the starlight contribution from the inner region of the host galaxy;
v) the cosmic background radiation.
All these contributions are evaluated in the blob comoving frame, where we calculate the 
corresponding inverse Compton radiation from all these contributions, and then transform
into the observer frame.

We calculate the energy distribution $N(\gamma)$ [cm$^{-3}$]
of the emitting particles at the particular time $R/c$, when the injection process ends. 
Our numerical code solves the continuity equation which includes injection, 
radiative cooling and $e^\pm$ pair production and reprocessing. 
Our is not a time dependent code: we give a ``snapshot" of the 
predicted SED at the time $R/c$, when the particle distribution $N(\gamma)$ 
and consequently the produced flux are at their maximum.

Since we are dealing with very powerful sources, the radiative cooling time of the 
particles is short, shorter than $R/c$ even for the least energetic particles.
This implies that, at lower energies, the $N(\gamma)$ distribution is proportional
to $\gamma^{-2}$, while, above $\gamma_{\rm b}$, $N(\gamma)\propto \gamma^{-(s_2+1)}$.
The electrons emitting most of the observed radiation have energies 
$\gamma_{\rm peak}$ which is close to $\gamma_{\rm b}$ 
(but these two energies are not exactly equal, 
due to the curved injected spectrum).

We model at the same time the
thermal disk (and IR torus) radiation and the non--thermal jet--emission.
The link between these two components is given by the amount of 
radiation energy density (as seen in the comoving frame of the emitting
blob) coming directly from the accretion disk or reprocessed by the BLR and
the IR torus.
This radiation energy density depends mainly on $R_{\rm diss}$, but
not on the adopted accretion rate or black hole mass
(they are in any case chosen to reproduce the observed thermal disk 
luminosity).

To calculate the flux produced by the accretion disk, we adopt a standard 
Shakura \& Sunyaev (1973) disk (see Ghisellini \& Tavecchio 2009).
This allow us to fit also the thermal radiation seen in the optical--UV range, and
to estimate the accretion rate and the black hole mass, as discussed below.

\subsection{Estimate of the black hole mass}

For all our sources, we interpret the near--IR, optical and UV
emission as due to the accretion disk.
For simplicity, we assume that the disk is an optically thin,
geometrically thick, Shakura--Sunjaev (1973) disk,
emitting a black-body at each radius.
The maximum temperature (and hence the $\nu F_\nu$ peak of 
the disk luminosity)
occurs at $\sim$5 \sc\ radii and scales as
$T_{\rm max}\propto (L_{\rm d}/L_{\rm Edd})^{1/4}M^{-1/4}$.
The total optical--UV flux gives $L_{\rm d}=\eta \dot M c^2$.
Once we specify the efficiency $\eta$,
we can derive both the black hole mass and the accretion rate.
Assuming a Schwarzschild black hole, we set $\eta=0.08$.

In Fig. {\ref{2149disk} we show a zoom of the SED of PKS 2149--306,
to illustrate the uncertainties on the estimated value of the mass of the central
black hole. 
The three SEDs shown correspond to $M=3\times 10^9$, $5\times 10^9$ and
$10^{10}$ solar masses. 
The maximum temperature of the accretion disk is a (albeit weak)
function of the black hole mass: the overall disk
emission becomes bluer for smaller black hole masses.
From Fig. \ref{2149disk} one can see that the $M=3\times 10^9 M_\odot$ case
gives a poor fit, while the $M=10^{10} M_\odot$ case tends to overproduce
the flux at lower optical frequencies.
We can conclude that, within our assumptions and  
when the data show the peak of the thermal emission,
the resulting estimate of the black hole mass is rather accurate
with uncertainties that are significantly smaller than a factor 2.

Three caveats are in order.
The first is that we are assuming a standard Shakura \& Sunjaev (1973)
disk, namely a disk geometrically thin and optically thick,
emitting black--body radiation at each annulus according to a standard
temperature profile (e.g. Frank, King \& Raine 2002).
For very large accretion rates, close to Eddington, the disk
structure might be modified, and in the inner region a funnel
may develop, as in the so called ``thick" and ``slim" disks
(see e.g. Abramowicz et al. 1988; Madau 1988; Szuszkiewicz, Malkan, \& Abramowicz 1996)
We have discussed this possibility for S5 0014+813 (Ghisellini et al. 2009).
If a funnel is present, the emitted disk radiation is not isotropic any longer,
and face--on observers (as for blazars)
would see an amplified radiation, and this would lead
to overestimate the black hole mass.
We will discuss again this point later.

The second caveat concerns the assumption of a \sc, non 
rotating black hole.
The efficiency $\eta$ for a Kerr hole and corotating
accretion disk is larger, as are the 
temperatures of the innermost radii.
But also gravitational light bending and redshift are larger,
and the calculation of the observed spectrum is not
as straightforward as in the \sc\ case (see e.g. Li et al. 2005).
Comparing the spectra calculated in the \sc\ and maximal Kerr case, 
for equal mass and accretion rate, and assuming that the disk emits 
black--body spectra at all radii, we have that the overall disk emission 
in the Kerr case is bluer and stronger above the emission peak, 
and very similar below (smaller frequencies are emitted at larger radii,
where the emitted flux is insensitive to the black hole rotation).
Decreasing the accretion rate would decrease the emission
peak (because the maximum temperature would be smaller),
but also the overall flux, both below and beyond the 
emission peak. 
Since we end up having a deficit in the low frequency part of the spectrum,
we must change also the mass, in the direction of an {\it increase} of it,
because this allows to have a smaller maximum temperature.
Therefore, to recover the original SED in the case of a Kerr hole,
we must decrease the accretion rate and increase the mass.
So assuming Kerr holes would not help in decreasing the derived masses.

The third caveat is that the assumption of a black--body spectrum 
may be too simplistic: for some regions of the disk 
a modified black--body may be a better choice.
Since a modified black--body is a less efficient radiator
than a pure black--body, the resulting spectrum, for the same accretion rate
and black hole mass, will be less powerful and bluer (because the disk
will be hotter).
If this occurs, then our derived black hole masses should be considered as 
lower limits.

\begin{figure}
\vskip -0.6 cm \hskip -0.3 cm
\psfig{figure=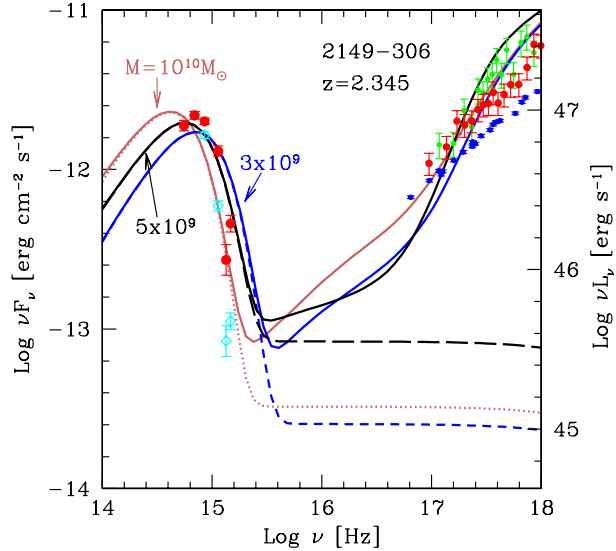,width=8.7cm,height=8.7cm}
\vskip -0.5 cm
\caption{
Zoom of the SED of PKS 2149--306, to show how the theoretical SED
changes by changing the black hole mass.
Three values are shown: $M=3\times 10^9$, $5\times 10^9$ and
$10^{10}$ solar masses.
Dotted, short and long dashed lines show the contribution of the accretion disk
and its X--ray corona, while solid lines show the sum of the thermal and the beamed
non-thermal components.
}
\label{2149disk}
\end{figure}

\begin{table*} 
\centering
\begin{tabular}{lllllllllllll}
\hline
\hline
Name   &$z$ &$R_{\rm diss}$ &$M$ &$R_{\rm BLR}$ &$P^\prime_{\rm i}$ &$L_{\rm d}$ &$B$ &$\Gamma$ 
    &$\gamma_{\rm b}$ &$\gamma_{\rm max}$ &$s_1$  &$s_2$  \\
~[1]      &[2] &[3] &[4] &[5] &[6] &[7] &[8] &[9] &[10] &[11] &[12] &[13]  \\
\hline   
0014+813    &3.366 &9.6e3  (800)  &4e10  &5.5e3  &0.12  &3e3  (0.5)  &0.33 &16    &150 &2e3 &--1   &3   \\ 
0222+185    &2.69  &660    (550)  &4e9   &1.45e3 &0.08  &210  (0.35) &2.94 &13.5  &40  &2e3 &0     &3.2 \\
0537--286   &3.104 &420    (700)  &2e9   &735    &0.13  &54   (0.18) &1.92 &15    &50  &2e3 &--1   &3   \\ 
0746+254    &2.979 &1.65e3 (1.1e3)&5e9   &866    &0.24  &75   (0.1)  &0.1  &15    &200 &5e3 &0.75  &2.6 \\
0805+614    &3.033 &270    (600)  &1.5e9 &581    &0.15  &34   (0.15) &2.54 &14    &60  &3e3 &--0.5 &3   \\  
0836+710    &2.172 &540    (600)  &3e9   &1.5e3  &0.22  &225  (0.5)  &3.28 &14    &90  &2e3 &--1   &3.6 \\
1210+330    &2.502 &420    (1.4e3)&1e9   &866    &0.08  &75   (0.5)  &0.73 &16    &80  &3e3 &--0.5 &3.5 \\
2126--158   &3.268 &1.8e3  (600)  &1e10  &2.7e3  &0.13  &750  (0.5)  &2.61 &14.1  &40  &2e3 &0     &3.3 \\ 
2149--306   &2.345 &1.2e3  (800)  &5e9   &1.24e3 &0.18  &150  (0.2)  &1.12 &15    &60  &3e3 &0     &3.3 \\ 
225155+2217 &3.668 &300    (1e3)  &1e9   &1.06e3 &0.15  &112  (0.75) &2.78 &15    &70  &2e3 &0     &4   \\
\hline
average     &3     &1.8e3  (1e3) &6e9   &1.3e3 &0.15  &180 (0.2)   &0.82 &15   &100 &2e3 &0     &3 \\ 
$\langle$LAT $z>2\rangle$ &2.2 &630    (700) &3e9   &948   &0.1   &90  (0.2)   &1.65 &15   &300 &4e3 &0     &2.8  \\
\hline
\hline 
\end{tabular}
\vskip 0.4 true cm
\caption{List of parameters used to construct the theoretical SED.
Col. [1]: name;
Col. [2]: redshift;
Col. [3]: dissipation radius in units of $10^{15}$ cm and (in parenthesis) in units of \sc\ radii;
Col. [4]: black hole mass in solar masses;
Col. [5]: size of the BLR in units of $10^{15}$ cm;
Col. [6]: power injected in the blob calculated in the comoving frame, in units of $10^{45}$ erg s$^{-1}$; 
Col. [7]: accretion disk luminosity in units of $10^{45}$ erg s$^{-1}$ and
        (in parenthesis) in units of $L_{\rm Edd}$;
Col. [8]: magnetic field in Gauss;
Col. [9]: bulk Lorentz factor at $R_{\rm diss}$;
Col. [10] and [11]: break and maximum random Lorentz factors of the injected electrons;
Col. [12] and [13]: slopes of the injected electron distribution [$Q(\gamma)$] below and above $\gamma_{\rm b}$;
The total X--ray corona luminosity is assumed to be in the range 10--30 per cent of $L_{\rm d}$.
Its spectral shape is assumed to be always $\propto \nu^{-1} \exp(-h\nu/150~{\rm keV})$.
The viewing angle $\theta_{\rm v}$ is $3^\circ$ for all sources.
}
\label{para}
\end{table*}

\section{Results}

Table \ref{para} lists the parameters used to fit the SED of the 10 blazars.
We find that they are all distributed in a narrow range, as expected, because 
we are dealing with extremely powerful objects, characterised by 
similar (albeit extreme) properties.

The distance from the black hole, $R_{\rm diss}$, where dissipation occurs  and
most of the observed radiation is produced, ranges from 500 to $1.4\times 10^3$ \sc\ radii.
Three sources (0746+253, 1210+330 and marginally 2149--306) 
have $R_{\rm diss}>R_{\rm BLR}$, while for all the others the 
location of the dissipation region lies within the BLR.
  
The injected power in the form of relativistic electrons is very similar,
around $10^{44}$ erg s$^{-1}$ for all sources, and since the cooling is severe,
all of it is transformed in radiation.
The accretion disk luminosity is at the level of 0.1--0.4 $L_{\rm Edd}$,
even if it spans a large range in absolute units, between $3\times 10^{46}$ and 
$2\times 10^{48}$ erg s$^{-1}$, due to the range of the black hole masses.
The bulk Lorentz factor is between 13 and 18.

\begin{table} 
\centering
\begin{tabular}{lllll}
\hline
\hline
Name   &$\log P_{\rm r}$ &$\log P_{\rm B}$ &$\log P_{\rm e}$ &$\log P_{\rm p}$ \\
\hline   
0048--071   &46.42 &46.98 &45.73 &47.49  \\
0222+185    &46.11 &46.42 &45.27 &47.89  \\
0537--286   &46.43 &45.74 &45.52 &48.01  \\
0746+253    &46.53 &44.36 &46.28 &47.85  \\
0805+614    &46.42 &45.34 &45.64 &47.99  \\
0836+710    &46.60 &46.36 &45.54 &48.00  \\
1210+330    &46.30 &44.95 &45.27 &47.79  \\
2126--158   &46.38 &47.27 &45.12 &47.84  \\
2149--306   &46.59 &46.18 &45.25 &48.02  \\
225155+2217 &46.48 &45.77 &45.74 &48.19  \\
\hline
average     &46.43 &45.96 &45.54 &47.92  \\
$\langle$LAT $z>2\rangle$ &46.3 &46.0 &44.8 &47.2 \\
\hline
\hline 
\end{tabular}
\vskip 0.4 true cm
\caption{
Logarithm of the jet power in the form of radiation, Poynting flux,
bulk motion of electrons and protons (assuming one proton
per emitting electron). Powers are in erg s$^{-1}$.
}
\label{powers}
\end{table}

\begin{figure}
\vskip -0.6 cm
\psfig{figure=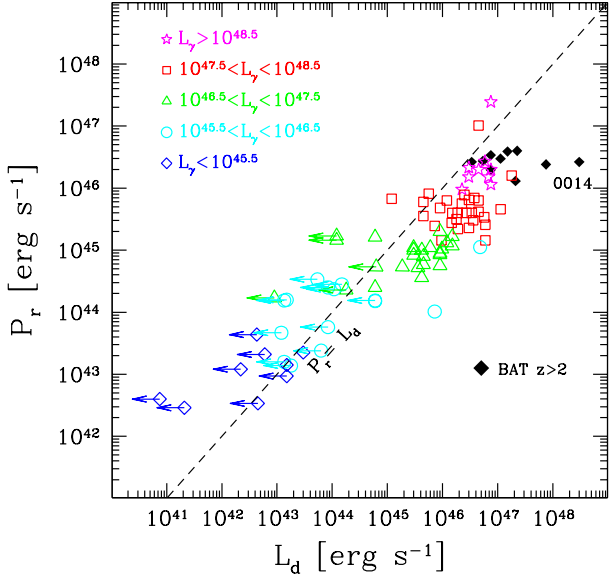,width=9cm,height=9cm}
\vskip -0.7 cm
\caption{The power spent by the jet to produce the radiation 
we see, $P_{\rm r}$, as a function
of the accretion disk luminosity $L_{\rm d}$.
BAT blazars (black diamonds) are compared with
the blazars in the {\it Fermi} 3--months catalogue of 
bright sources detected above 100 MeV. 
The latter have different symbols according to their
$\gamma$--luminosity, as labelled.
}
\label{lrld}
\end{figure}

\begin{figure}
\vskip -0.6 cm
\psfig{figure=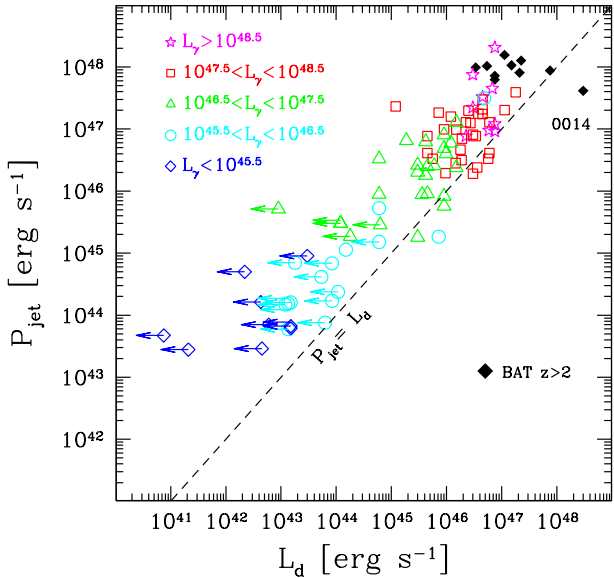,width=9cm,height=9cm}
\vskip -0.5 cm
\caption{The jet power $P_{\rm jet}$ as a function
of the accretion disk luminosity $L_{\rm d}$.
Symbols as in Fig. \ref{lrld}.
}
\label{pjld}
\end{figure}

\begin{figure}
\vskip -0.6 cm
\psfig{figure=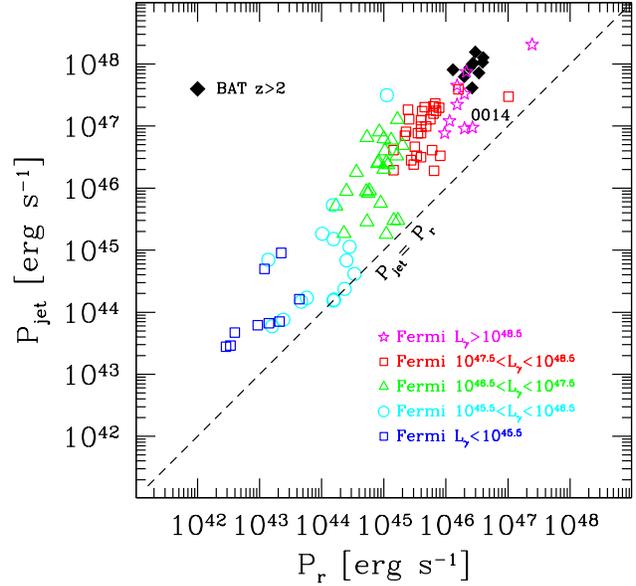,width=9cm,height=9cm}
\vskip -0.5 cm
\caption{The jet power $P_{\rm jet}$ as a function
of the power spent by the jet to produce the radiation 
we see, $P_{\rm r}$.
Symbols as in Fig. \ref{lrld}
}
\label{pjlr}
\end{figure}

\subsection{Comparison with bright Fermi/LAT  blazars}

It is instructive to compare the jet power and accretion disk luminosities of 
the BAT blazars of our sample with all the bright blazars detected by the 
LAT 3--months survey, as analysed in G10.
The jet carries power in the form of bulk motion of particles (electrons and
protons) and magnetic field.
Furthermore, we can calculate the power spent by the jet to produce
the radiation we see.
The different components of the jet power are
\begin{equation}
P_i  \, =\, \pi r_{\rm diss}^2 \Gamma^2\beta c \, U^\prime_i
\end{equation}
where $U^\prime_i$ is the energy density of the $i$ component,
as measured in the comoving frame.
We define the total jet power as the sum of the electron ($P_{\rm e}$), 
proton ($P_{\rm p}$), found assuming one proton per electron, and magnetic
field ($P_{\rm B}$) contributions, while we call $P_{\rm r}$ the
power spent in producing radiation. 
Table \ref{powers} lists the jet powers for the 10 BAT blazars.

\vskip 0.3 cm
\noindent
$P_{\rm r}$ vs $L_{\rm d}$ --- 
Fig. \ref{lrld} shows $P_{\rm r}$ as a function of the accretion
luminosity $L_{\rm d}$ for our BAT blazars, together with all
the LAT blazars in G10.
These include BL Lac objects for which only an upper limit to their disk luminosity
could be found (shown by arrows).
Black filled diamonds correspond to the BAT blazars.
The power 
$P_{\rm r} =\pi r_{\rm diss}^2 \Gamma^2\beta c \, U^\prime_{\rm rad}$,
can be re--written as [using $U^\prime_{\rm rad}=L^\prime/(4\pi r_{\rm diss}^2 c)$]:
\begin{equation}
P_{\rm r}  \, =\,  L^\prime {\Gamma^2 \over 4} \, =\, L {\Gamma^2 \over 4 \delta^4}
\, \sim \, L {1 \over 4 \delta^2}
\end{equation} 
where $L$ is the total observed non--thermal luminosity
($L^\prime$ is in the comoving frame) and $U^\prime_{\rm rad}$ is the 
radiation energy density produced by the jet (i.e.
excluding the external components).
The last equality assumes $\theta_{\rm v}\sim 1/\Gamma$.
This quantity is almost model--independent,
since it depends only on the adopted $\delta$, that can be estimated 
also by other means, namely superluminal motions.
Therefore Fig. \ref{lrld} shows two quantities that are (almost) 
model--independent.
The BAT blazars are the most extreme, lying at the upper end
of the $P_{\rm r}$--$L_{\rm d}$ distribution.
All follow the trend defined by the LAT FSRQs, with the exception of S5 0014+813,
that has an accretion disk luminosity larger than what expected from its jet 
luminosity.

\vskip 0.3 cm
\noindent
$P_{\rm jet}$ vs $L_{\rm d}$ ---
Fig. \ref{pjld} shows the jet power $P_{\rm jet}=P_{\rm p}+P_{\rm e}+P_{\rm B}$ 
as a function of the disk luminosity.
Again, the BAT blazars have the most powerful jets and disks,
and S5 0014+813 appears to be an outlier with respect to the general trend 
defined by the ensemble of LAT and BAT blazars.

\vskip 0.3 cm
\noindent
$P_{\rm jet}$ vs $P_{\rm r}$ ---
Fig. \ref{pjlr} shows the jet power as a function of $P_{\rm r}$,
the power in radiation.
It shows the efficiency of the jet in converting its bulk 
power into radiation.
Low power BL Lacs are the most efficient in converting $P_{\rm jet}$ 
into $P_{\rm r}$, while powerful blazars are less efficient.
The BAT blazars follows this trend, and in this plot S5 0014+813 is not an outlier.

This suggests that S5 0014+813 indeed has an over--luminous accretion disk 
with respect to its jet, and this favours the hypothesis, discussed in Ghisellini
et al. (2009) that the thermal radiation from the inner parts of the disk
is geometrically collimated by the presence of a funnel, possibly as a consequence of the 
large accretion rate in Eddington units (likely to be even larger than what calculated
here and in Ghisellini et al. 2009, because we assumed a standard disk).

For the other BAT blazars, instead, there is no need to invoke a non--standard
accretion disk, since they follow the jet power -- disk luminosity relation 
defined by less extreme blazars, selected from the {\it Fermi}/LAT survey.
The latter have luminosities around 0.1 of the Eddington one or even less,
and should indeed have standard disks. 
This suggests that the black hole masses estimated for the other BAT blazars are
robust, and not a consequence of our assumption of a standard accretion disk.

\begin{figure}
\vskip -0.6 cm \hskip -0.4 cm
\psfig{figure=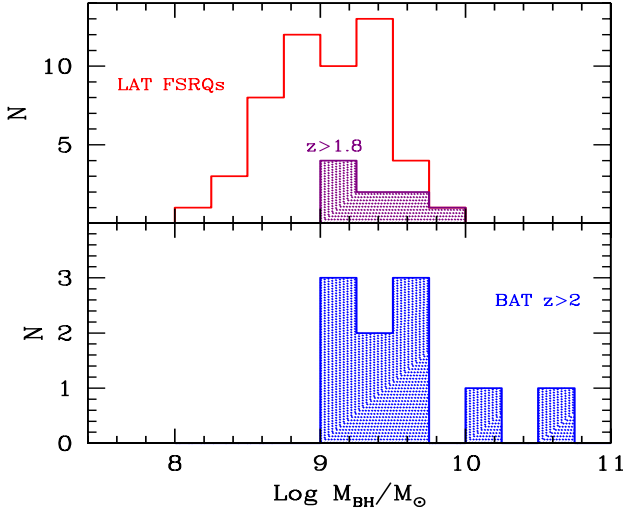,width=9.5cm,height=8.cm}
\vskip -0.5 cm
\caption{
Comparison of black hole mass distributions.
The top panel shows all {\it Fermi}/LAT FSRQs (i.e. excluding BL Lacs),
and the 9 {\it Swift}/BAT FSRQs with $z>1.8$.
The bottom panel shows the mass distribution of the 10 BAT 
blazars at $z>2$.
}
\label{masses}
\end{figure}

\begin{figure}
\vskip -0.6 cm \hskip -0.4 cm
\psfig{figure=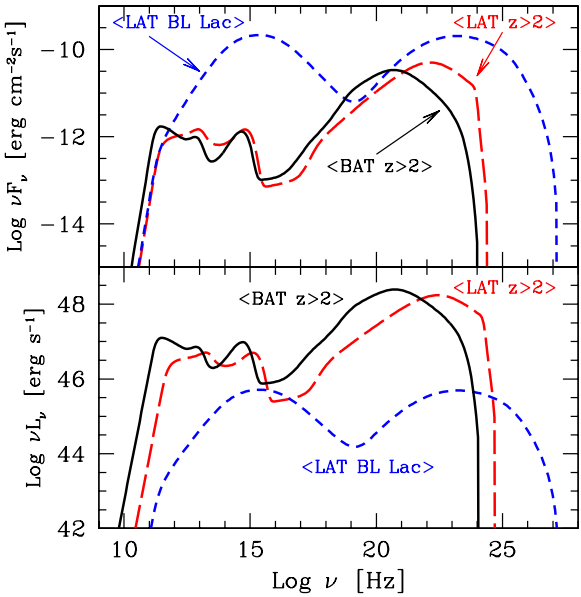,width=9.5cm} 
\vskip -0.8 cm
\caption{Average SEDs for the sources in our samples,
for all FSRQs detected by {\it Fermi} and for the subsample
of the 8 {\it Fermi}--blazars at redshift $z>1.8)$.
The top and the bottom panels show the fluxes and luminosities,
respectively.
The shown frequencies are calculated in the rest frame
of the source for the luminosity plot, and are the observed
ones for the flux plot. 
In Tab. \ref{para} we list the parameters used to construct
the shown SEDs for FSRQs, while, for the LAT BL Lacs, we report the same
SED as shown in G10.
}
\label{average}
\end{figure}

It is then interesting to compare the black hole masses derived for our
BAT blazars with those of the 3--months survey of LAT, and then
to the sub--sample (of 9 objects) of LAT blazars at $z>1.8$.
This is done in Fig. \ref{masses}: 
high redshift BAT blazars have a black hole mass distribution extending 
to larger values than the high--$z$ LAT blazars.
The numbers are too small for a statistical comparison, and a K--S
test returns a large probability that the two distributions
are drawn from the same one.
In any case, we can safely conclude that the hard X--ray 
selection is a very efficient
way to pinpoint the most powerful extreme of the blazar population, possibly
more than selecting them through their {\it Fermi}/LAT $\gamma$--ray flux.
We can wonder if this merely depends on the sensitivities
of the current BAT and LAT instruments, or if indeed the most
powerful blazar jets are more luminous in hard X--rays than in 
the 0.1--10 GeV band.
In Fig. \ref{average} we try to answer to this question 
by showing three theoretical
SEDs corresponding to; i) the average SED of BL Lacs detected by the 3--months
LAT survey (see the used parameters in G10); 
ii) the average SED of the $z>2$ FSRQs detected by the same
LAT survey, and iii) the average SED of our $z>2$ BAT FSRQs.
For the latter two SEDs we used the parameters listed at
the end of Tab. \ref{para}.
We show both the flux (top panel) and the luminosity (bottom) SEDs.
The shown frequencies are calculated in the rest frame
of the source for the luminosity plot, and are the observed
ones for the flux plot. 
Fig. \ref{average} shows that:
\begin{itemize}
\item 
LAT BL Lacs and high redshift FSRQs
have the same average flux in hard X--rays;
\item 
LAT BL Lacs are on average brighter than high--$z$ FSRQs
in $\gamma$--rays;
\item 
Comparing high redshift LAT and BAT FSRQs, we see that the latter are 
fainter in $\gamma$--rays and in fact none of our 10 BAT blazars 
has been detected in the 3--months LAT survey;
\item 
In the luminosity plot, LAT BL Lacs are the least luminous
(most of them are at $z<0.5$, see Fig. \ref{iz});
\item 
Comparing high--$z$ BAT and LAT FSRQs, we
see that BAT blazars are slightly more luminous (in bolometric
terms) even if they are less powerful $\gamma$--ray sources.
In the entire X--ray band they are more powerful than LAT blazars.

\end{itemize}

We have checked if some of the high redshift LAT and BAT blazars
are present on the SDSS survey,
with their black hole mass estimated through the FWHM of the emission
lines and the ionising continuum luminosity. 
In the compilation of Shen et al. (2008) we have found 
RGB J0920+446 ($z=2.190$) with different mass estimates, ranging from 
$M=2\times 10^9 M_\odot$ and $5\times 10^{10} M_\odot$, to be compared with
our estimate of $M= 6\times 10^9M_\odot$ (Ghisellini, Tavecchio \& Ghirlanda 2009); 
4C +38.41 (=1633+382, $z=1.814$) with a mass in the range $(1.6-8)\times 10^9 M_\odot$ 
(our estimate: $M=5\times 10^9 M_\odot$, Ghisellini, Tavecchio \& Ghirlanda 2009)
and J074625.87+2549 with  $M=6\times 10^9 M_\odot$ (our estimate: $M=7\times 10^9 M_\odot$,
see Tab. \ref{para}).
%
%
Also PKS 1502+106 ($z=1.839$) is present in the SDSS (DR7) 
catalogue, but with no mass estimate yet.

All this phenomenology can be understood in simple terms on the basis
of the blazar sequence: the most powerful blazars
are FSRQs whose synchrotron peak is located in the sub--mm band, and 
whose high energy peak is located in the MeV band.
So the 0.1--10 GeV luminosity, being in a band relatively far from the peak, 
is not very large.
On the contrary, instruments sensitive in the hard X--ray band
can catch these sources at their emission peak, making 
this band the optimal one in finding the upper end of the blazar
luminosity distribution.

\section{Discussion}

One of our major result is that {\it all} the 10 BAT blazars
studied here have a black hole heavier than $10^9$ solar masses.
Among the BAT blazars they are the most luminous 
(all of them have $L_X>2\times 10^{47}$ erg s$^{-1}$) and 
the most distant (all of them have $z>2$).
Since these objects are at high redshifts, our finding has
important implications on the number density of heavy black holes,
especially if we consider that for each blazar pointing at us, there
must be hundreds of similar sources (having black holes of similar masses)
pointing elsewhere.
Using a mix of simple theoretical and observational considerations 
we will find in the following what we consider a conservative 
``minimal mass function" for heavy black holes associated to 
radio--loud objects.
But before doing this we first discuss the implications of our results for 
future hard X--ray missions.

\subsection{Implications for future missions}
\label{future}

{\bf NHXM --} The {\it New Hard X--ray Mission 
(NHXM)}\footnote{http://www.brera.inaf.it/NHXM2/}
is a project for a satellite hosting 4 mirror modules able to concentrate
X--ray photons in the 0.2--80 keV range (Pareschi et al. 2009).
This is achieved in part by a long (10 meter) focal length and partly by a
multi--layer coating.
A 2--30 keV polarimeter is foreseen at the focus of one of the mirror modules.
The angular resolution is better than $20^{\prime\prime}$ at 30 keV.
The sensitivity above 10 keV is 2--3 orders of magnitude better than anything 
already flown.
On the other hand, the long focal length limits the field of view to $\sim 12^\prime\times 12'$.

The main contribution of this mission to the issue of finding large black hole masses
at high redshift is to observe pre--selected candidates in the hard X--ray region
of the spectrum, and consequently to assess the non--thermal nature of this
emission. 
By establishing a large jet power, it will hint to a corresponding
large accretion rate and thus a large black hole mass.
In this context, we stress that the blazars we have analysed in this paper are
all strong ($>0.1$ Jy) radio sources and very hard in the ``classical" 2--10 keV band.
These properties can then be taken as selection criteria
to construct a useful sample of radio--loud sources to be observed by ${\it NHXM}$.

\vskip 0.3cm

\noindent
{\bf EXIST --} The {\it Energetic X--ray Imaging Survey Telescope 
(EXIST)}\footnote{http://exist.gsfc.nasa.gov/}
is a proposed {\it Medium Class Mission} to conduct the most sensitive 
full--sky survey for black holes on all scales (from the stellar to the supermassive ones).
To this purpose {\it EXIST} has been specifically designed to have   
onboard three complementary instruments:
\begin{itemize}
\item 
a large area (4.5 m$^2$), wide--field ($70^\circ \times 90^\circ$) hard X-ray 
(5--600 keV) imaging (2 arcmin resolution and $\sim 20^{\prime\prime}$ 
localisation for  $5\sigma$ sources) coded mask telescope (HET, High Energy Telescope);
\item
a soft X-ray imager (SXI, focusing telescope with CCD, $\sim 2^{\prime\prime}$ 
localisations) 
operating over the 0.1--10 keV energy range with {\it XMM--Newton}--like (1--telescope) area; 
\item
an optical/NIR telescope (IRT) with 1.1 m diameter with instruments 
covering the wavelength range 3000--22000 \AA $\ $ with R$\sim$3000
spectrograph and $4^{\prime} \times 4^{\prime}$ imaging array with objective 
prism capability.
\end{itemize}

HET has approximately a 20 times better sensitivity than BAT, and extends
the energy range to lower and higher energies: it can then see powerful and distant
blazars {\it at or very close to their peak of emission}.
Furthermore, the operation mode (similar to {\it Fermi}), patrolling the entire
sky every three hours, is ideal to discover rare objects.  
SXI can give crucial information about the level and shape of the 
0.2--10 keV spectrum as well as to locate the X-ray sources on sky with an accuracy 
of few arcsec.
Finally, IRT can take the IR and optical spectra (and/or photometry) of these rare sources,
and can catch powerful blazars {\it where their accretion disk peaks}.
Given the expected flux levels of IR--optical and X--ray fluxes of high redshift 
powerful blazars, these ``follow--up" observations will be very inexpensive.
{\it EXIST} has been designed to break the ``multi-wavelength investigation 
bottleneck" by having on--board a suite of instruments on unprecedented wavelength 
coverage that operate simultaneously. 

To illustrate these concepts, Fig. \ref{z8} shows the SED of PKS 2149--306,
the theoretical fitting model, and the same model for a source located at $z=8$
having the same total luminosity.
We superimpose also the expected sensitivities of the three {\it EXIST}
instruments (for the indicated exposure times).
If blazars like PKS 2149--306 exist at high redshifts, {\it EXIST} will be able to
find them and to characterise their general physical
properties in a rather complete way.
These include the jet power, the accretion disk luminosity and
the black hole mass, that can be estimated in the same way as done in this paper. 

\begin{figure}
\vskip -0.6 cm \hskip -0.4 cm
\psfig{figure=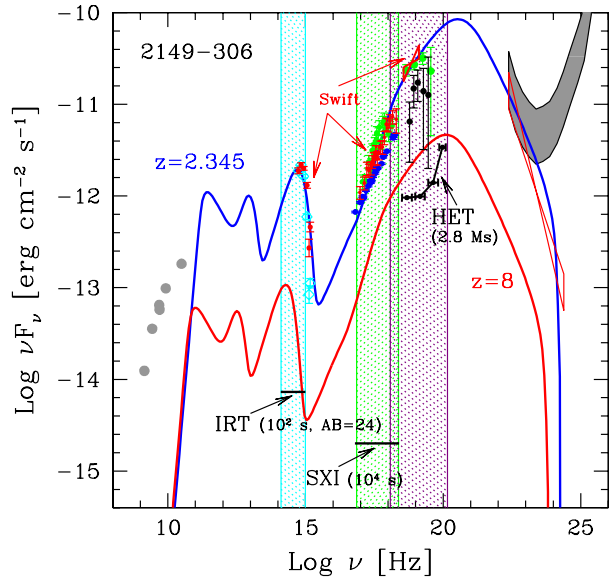,width=9cm} 
\vskip -0.8 cm
\caption{\
The SEDs and model for PKS 2149--306 at its actual redshift ($z=2.345$)
and what the model SED would appear if the source were at $z=8$.
We also show the limiting sensitivities for the three instruments foreseen
to be onboard the {\it EXIST} mission: the high energy coded mask
(HET, sensitive in the 5--600 keV range), the soft X--ray Telescope (SXI,
operating in the 0.1--10 keV energy range), and the IR--optical telescope (IRT).
We indicate the exposure time needed to reach the shown sensitivities.
It is remarkable that powerful blazars like PKS 2149--306, if the existed 
even at $z=8$, could easily be detected by hard X--ray telescopes like {\it EXIST}/HET.
The presence of SXI and especially the IR--optical telescope would also
allow to find the redshift and to very easily provide a complete spectrum of the 
accretion disk at its emission peak, thus yielding a robust estimate
of the accretion rate and the black hole mass.
}
\label{z8}
\end{figure}

\begin{figure}
\vskip -0.6 cm  
\psfig{figure=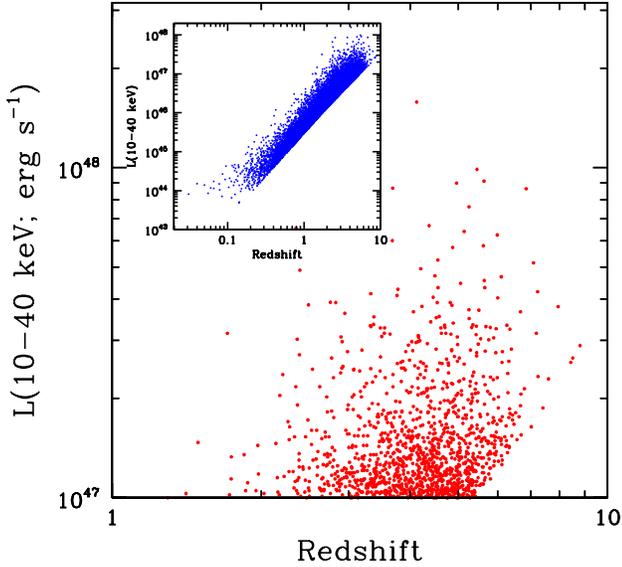,width=9cm} 
\vskip -0.8 cm
\caption{
X--ray luminosity as a function of redshift 
for a simulated sample of blazars detectable by the two year all 
sky survey of {\it EXIST}. 
The inset shows the entire distribution, while the main figure 
zooms the region of high luminosity. 
This expected sample of blazars detectable by 
{\it EXIST} has been produced by extrapolating the cosmological evolution 
properties of the BAT blazars as derived by A09.
}
\label{rdc1}
\end{figure}

\begin{figure}
\vskip -0.6 cm  \hskip 0.4 cm
\psfig{figure=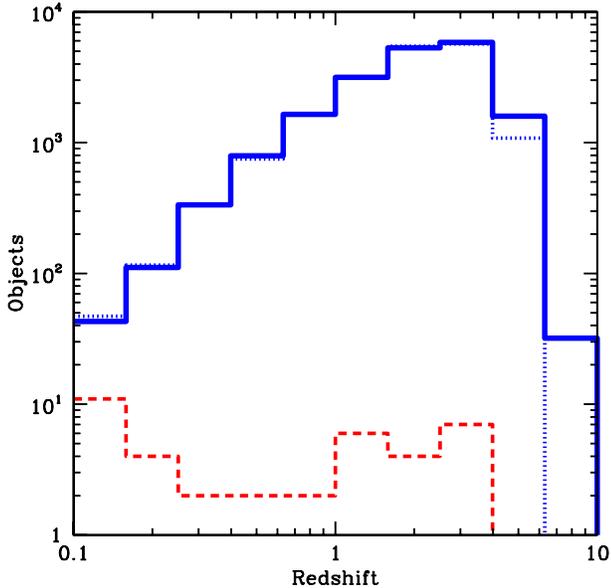,width=8.3cm} 
\vskip -0.2 cm
\caption{
The redshift distribution of the blazars detectable by 
the 2 year all sky survey of {\it EXIST}
compared with the redshift distribution of the 
BAT blazars (dashed line).
The dotted histogram correspond to the ``minimal" LF as
discussed in the text.
}
\label{rdc2}
\end{figure}

Of course the choice of $z=8$ (for Fig. \ref{z8}) does not imply that blazars
of this kind do exist at this redshift.
To evaluate the number of blazars detectable by {\it EXIST} in the all sky survey we 
used the X--ray luminosity function and the cosmological evolution model of blazars as 
recently derived by A09 from the analysis of the BAT data.
In particular, we used the modified pure luminosity evolution model with
best fit parameters as reported in figure 9 of 
A09\footnote{We used $\gamma_1=-0.87$, $\gamma_2=2.73$, 
$L_{\star}=1.8\times 10^{44}$ erg s$^{-1}$, 
$k=4.8$, $\gamma=-0.36$, see A09 for details.}, 
re--normalising the luminosity function to the number of blazars detected by BAT (38 sources).
A power--law spectrum with photon index equal to 1.5 has been assumed to convert 
the (15--55 keV) A09
luminosity function to the 10--40 keV energy range and to 
compute k-corrections.

Assuming a flux limit in the 10--40 keV energy band of $\sim 8\times 10^{-13}$ 
erg cm$^{-2}$ s$^{-1}$ for the 2--years {\it EXIST} survey data, 
we expect $\sim 19,000$ blazars in the all sky survey.
Their distribution in the luminosity--redshift
plane is shown in Fig. \ref{rdc1}, while their redshift distribution (compared with 
the redshift distribution of the BAT blazars) is shown in Fig. \ref{rdc2} (solid line).
Table \ref{numbers} list the expected number of blazars as a function of $z$,
together with the number of high X--ray luminosity blazars
possibly hosting a heavy black hole (columns 2 and 3, respectively).
This makes it clear the great discovery potential of {\it EXIST}; 
more than 1500 blazars are expected at redshift greater than 4 
and about a dozen at redshift greater than 7.

The expected number of high X--ray luminosity blazars
reported in column 3 of Table \ref{numbers} clearly depends  
on the extrapolation of the A09 cosmological 
evolution model at redshift greater than $\sim 4$, a redshift range 
where no BAT data are available. 
To derive a ``strict" lower limit to the number of  high X--ray luminosity 
blazars detectable by {\it EXIST} we have used a cosmological 
evolution model that is equal to the A09 one up to $z\sim 4.3$ 
(where they measure the 
peak of the density of high X--ray luminosity blazars) but having 
an exponential cutoff beyond $z=4.3$
[$L_z = L_{z=4.3} \times \exp(z-4.3)$ for $z>4.3]$. 
This LF is consistent with the lower limits on the number
of blazars known between $z=4$ and $z=5$ (see Fig. \ref{BHhighz}).
In the following we will call this evolution model ``minimal LF".
The expected number of blazars computed using this  ``minimal LF"
are reported in Table \ref{numbers} (column 4 and 5), while their 
redshift distribution is shown in  Fig. \ref{rdc2} (dotted line). 
As expected the only differences are at redshift greater than 4, where
for this rather extreme ``minimal LF" we expect to detect about 60 
high X--ray luminosity blazars at $z>4$, 5 at $z>5$ and none at $z>6$.
The real number of high X--ray luminosity blazars detectable by {\it EXIST} 
are probably in--between the numbers reported in Table \ref{numbers} (columns 3 and 5).


\begin{table} 
\centering
\begin{tabular}{lllll}
\hline
\hline
$z$  &$N_{\rm A09}(>z)$  &$L_X>L_{\rm thr}$ &$N_{\rm min}(>z)$ &$L_X>L_{\rm thr}$  \\
\hline
0  &1.9e4 &223 &1.83e4 &121   \\
1  &1.6e4 &223 &1.53e4 &121   \\ 
2  &1.0e4 &222 &9.7e3  &120   \\
3  &4.9e3 &199 &4.3e3  &102   \\
4  &1.6e3 &154 &1.0e3  &57    \\
5  &335   &76  &41     &5   \\
6  &58    &24  &1      &0   \\
7  &11    &9   &0      &0   \\
8  &3     &3   &0      &0   \\
\hline
\hline 
\end{tabular}
\vskip 0.4 true cm
\caption{
Number of blazars detectable by {\it EXIST}
above a given redshift, according to the luminosity function 
derived in A09 and the limiting sensitivity of {\it EXIST}
for the two year all sky survey. 
The third column gives the number of objects, above
a given $z$ and with X--ray luminosities above 
$L_{\rm thr}= 2\times 10^{47}$ erg s$^{-1}$.
The fourth and fifth column give the total number of blazars,
and those exceeding $L_{\rm thr}$, for the ``minimal" LF
discussed in the text.
}
\label{numbers}
\end{table}

\subsection{Black hole--dark halo connection at high redshifts}

It is well known that growing a billion solar masses black hole within 
a billion years from the Big Bang is a challenge
for hierarchical models of structure formation (e.g., Haiman 2004; 
Shapiro 2005; Volonteri \& Rees 2005; Volonteri \& Rees 2006; Tanaka 2009). 
Assuming accretion at the Eddington rate, a black hole mass 
increases in time as:
\beq 
M(t)=M(0)\,\exp\left(\frac{1-\epsilon}{\epsilon}\frac{t}{t_{\rm Edd}}\right),
\eeq 
where $t_{\rm Edd}=0.45\, {\rm Gyr}$ and $\epsilon$ is the radiative efficiency. 
For a ``standard" radiative efficiency $\epsilon\approx0.1$, and a seed mass  
$M(0)=10^2-10^4 M_\odot$, it takes at least 0.7--0.9 Gyr to 
grow up to $\simeq 10^{10} M_\odot$. 
The cosmic time at $z=4$ is 1.5 Gyr, 
but it is only 0.9 Gyr at $z=6$, and 0.7 Gyr at $z=7$. 
We expect therefore that billion solar masses black holes at 
higher and higher redshift becomes increasingly rare.

We will provide an estimate of the number density of black 
holes with $M>10^9 M_\odot$ as a function of redshift that is 
{\it independent} of the formation and growth efficiency of black holes. 
Empirical correlations have been found between the black hole mass 
($M$) and the central stellar velocity dispersion ($\sigma$) of the 
host (G\"ultekin et al. 2009 and references therein), and between 
the central stellar velocity dispersion and the asymptotic circular 
velocity ($V_{\rm c}$) of galaxies (Ferrarese 2002; Pizzella et al. 2005; 
Baes et al. 2003). 
\beq
\sigma=200 \kmps \left(\frac{V_{\rm c}}{320 \kmps}\right)^{1.35}
\eeq
and
\beq
\sigma=200 \kmps \left(\frac{V_{\rm c}}{339 \kmps}\right)^{1.04}
\eeq
as suggested by Pizzella et al. (2005) and Baes et al. (2003), respectively.


The latter is a measure of the total mass of the dark matter halo of 
the host galaxies. 
A halo of mass $M_{\rm h}$ collapsing at redshift $z$ has a circular velocity
\beq V_{\rm c}= 142 \kmps \left[\frac{M_{\rm h}}{10^{12} \ M_{\sun} }\right]^{1/3} 
\left[\frac {\Omm}{\Ommz}\ \frac{\Delta_{\rm c}} {18\pi^2}\right]^{1/6} 
(1+z)^{1/2}  
\eeq 
where $\Delta_{\rm c}$ is the over--density at virialization relative 
to the critical density. 
For a WMAP5 cosmology we adopt here the  fitting
formula (Bryan \& Norman 1998) $\Delta_{\rm c}=18\pi^2+82 d-39 d^2$, 
where $d\equiv \Ommz-1$ is evaluated at the collapse redshift, so
that $ \Ommz={\Omm (1+z)^3}/({\Omm (1+z)^3+\Oml+\Omk (1+z)^2})$. 

We will further assume that the black hole--$\sigma$ scaling is:
\beq
M=10^9 M_\odot \left(\frac{\sigma}{356 \kmps} \right)^4.
\eeq
and that these scaling relations observed in the local universe 
hold at all redshifts.  
Therefore we derive the relationship 
between black hole and dark matter halo mass (see also Bandara et al. 2009):
\beq
M_{\rm h}=4.1\times10^{13} M_\odot  \left[\frac{M}{10^9 M_\odot} \right]^{0.56}
\left[ \frac{\Omm}{\Ommz} \frac{\Delta_c} {18\pi^2} \right]^{-1/2} 
(1+z) ^{-3/2} 
\eeq
and
\beq
M_{\rm h}=7.1\times10^{13} M_\odot \left[\frac{M}{10^9 M_\odot} \right]^{0.72}
\left[ \frac{\Omm}{\Ommz} \frac{\Delta_{\rm c}} {18\pi^2} \right]^{-1/2} 
(1+z)^ {-3/2} 
\eeq
%

\begin{figure}
\vskip -0.6 cm \hskip -0.3 cm
\psfig{figure=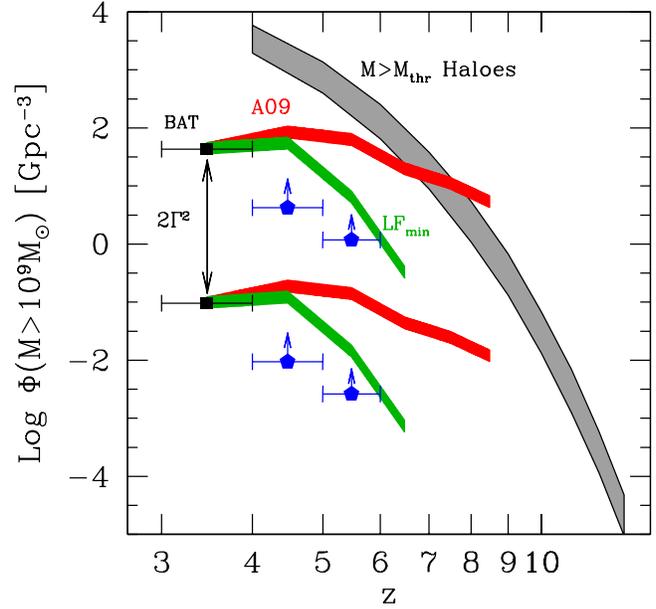,width=9.5cm} 
\vskip -0.8 cm
\caption{
Number density of black holes with $M>10^9 M_\odot$ as a function of redshift. 
The grey stripe is based purely on connecting black hole mass to halo mass
(upper envelope: Pizzella et al. 2005; bottom: Baes et al. 2003) 
and estimating the number density of black holes via the number density of their 
host dark matter halos using the Press \& Schecter formalism (see text).
In the lower part of the figure we report, as grey stripes (red and green in the
electronic version), the mass function
$\Phi(z,M>10^9M_\odot)$ for blazars as derived from the luminosity function of A09, 
considering $L_X>2\times 10^{47}$ erg s$^{-1}$. 
These stripes correspond two ``extreme" cases 
of cosmological evolution for the blazars population beyond redshift 4: 
the upper (red) stripe corresponds to the 
cosmological evolution model of A09 extrapolated up to $z \sim 9$, 
while the lower stripe corresponds to its ``high--$z$ cut-off
version" as discussed in the text.
The filled square in the  $3<z<4$ bin is taken directly from Fig. 10 of A09.
The filled pentagons and arrows are the lower limits derived from
the existence of a few blazars in the 2 redshift bins.
In the upper part of the figure we show the same points/stripes
up--shifted by the factor $2\Gamma^2=450$, to account for misaligned sources.
In this way the upper stripe of the mass function
$\Phi(z,M>10^9M_\odot)$ for radio--loud sources is in conflict,
at high $z$, with the estimates derived by massive halos, while
the lower (green) stripe derived through the ``minimal LF" is consistent.
}
\label{BHhighz}
\end{figure}

The number density of black holes with $M>10^9 M_\odot$, 
therefore corresponds to the number density of halos with mass 
$M_{\rm h}>M_{\rm thr}$, if $M_{\rm thr}$ is the mass of a halo that hosts a billion 
solar masses black hole.  
We estimate the number density of dark matter 
halos using the Press \& Schecter formalism (Sheth \& Tormen 1999).
In Fig. \ref{BHhighz} we show the results as the grey stripe,
encompassing the values found by the two methods above.
This estimate ignores any issue related to black hole 
formation and growth timescale. 
The grey stripe can then be considered as a rough upper limit to
the number density of heavy black holes as a function of redshift.

\subsection{Large black hole masses at high redshift}

These model predictions can be compared 
with the volume density of high redshift blazars hosting a 
black hole of mass larger than $10^9 M_\odot$.
The latter can be found using the cosmological 
evolution model of A09 along its high--$z$ cut--off (i.e. ``minimal")
version (see \S \ref{future}),
assuming, as we have found in this paper, that all blazars with $L_X>2\times 10^{47}$
erg s$^{-1}$ have a $M>10^9 M_\odot$ black hole.
We cannot exclude that blazars with lower X--ray luminosity also host
massive black holes, so the ``observational" points, strictly
speaking, are lower limits.

The lower part of Fig. \ref{BHhighz} shows two stripes (red and green,
in the electronic version) 
corresponding to two mass functions $\Phi(z, M>10^9 M_\odot)$ of blazars, 
both calculated for X--ray luminosities larger than $2\times 10^{47}$ erg s$^{-1}$.
The flatter (red) stripe corresponds to the 
cosmological evolution model of A09 extrapolated up to $z \sim 9$, 
while the steeper (green) stripe corresponds to its ``high z cut-off
version" (``minimal" LF, see \S \ref{future} for details).

The lower limits shown with a (blue) pentagon and arrow   
in the 4--5 redshift bin corresponds
to the existence of at least 4 blazars for which we have estimated a
black hole mass larger than $10^9 M_\odot$.
They are 
RXJ 1028.6--0844 ($z=4.276$; Yuan et al. 2005);
GB 1508+5714 ($z=4.3$; Hook et al. 1995);
PMN J0525--3343 ($z=4.41$; Worsley et al. 2004a) and
GB 1428+4217 ($z=4.72$; Worsley et al. 2004b).
There are other 3 blazars with $4<z<5$ discussed in Yuan et al. (2006),
but they have X--ray luminosities smaller than $10^{47}$ erg s$^{-1}$.
The other lower limit shown by the (blue) pentagon and arrow 
in the 5--6 redshift  bin corresponds to the existence of at least one blazar, 
Q0906+6930 at $z=5.47$, with an estimated
black hole mass of $2\times 10^9 M_\odot$ (Romani 2006).

All these points concerns sources pointing at us.
The real density of heavy black holes must account for the 
much larger population of misaligned sources.
We have then multiplied the mass function $\Phi(z, M>10^9M_\odot)$
of blazars and the other lower limits by 
$2\Gamma^2=450$, i.e. we have assumed an average $\Gamma$--factor of 15,
appropriate for the BAT blazars analysed here.
Fig. \ref{BHhighz} show the resulting points.

The mass function of heavy black holes of all jetted sources 
is now close or even greater (if we extend the cosmological 
evolution model of A09 beyond z$\sim$4) than
the upper limit defined by ``halo--black holes" 
(grey stripe) at the largest redshifts.
The mass function derived by the ``minimal" LF is instead
consistent.


To summarise: the BAT blazar survey allowed to meaningfully
construct the hard X--ray LF of blazars.
We have shown that its high luminosity end can be 
translated into the mass function of black holes with more
than one billion solar masses.
Up to $z=4$, where we do see blazars, the cosmological evolution model, 
as derived by A09, is secure.
Beyond $z=4$ it depends strongly on the assumed evolution.
We have then constructed the minimal evolution
consistent with the existing data and the (few)
existing lower limits.
As Fig. \ref{BHhighz} shows, the true mass function of heavy black
hole in jetted sources should be bracketed by the two shown
mass functions derived from the A09 and the ``minimal" LF.
The true mass density should then lie in--between the two
possible choices.
The implications of this finding are far--reaching, and
we plan to investigate them in a forthcoming study 
(Volonteri et al. in preparation).


\section{Conclusions}

We summarise here our main conclusions.

\begin{itemize}

\item In the $\alpha_X$--$L_X$ plane the ensemble of blazars
detected by BAT separate quite clearly in BL Lac objects and
FSRQs: the former have steeper spectra and lower luminosities.
This is a manifestation of the blazar sequence, since low power blazars
are characterised by a population of emitting electrons with a large
energy break, implying larger synchrotron and Inverse Compton frequency peaks.
In the hard X--ray range we often see, in BL Lacs, the steep synchrotron
tail of emission, while in more powerful blazars (i.e. FSRQs), we see
the hard Inverse Compton component.

\item In the same plane there is an indication of a ``divide", namely
BL Lacs and FSRQs separate in luminosity, at a few times $10^{45}$
erg s$^{-1}$.
This behaviour mirrors what occurs for {\it FERMI}/LAT blazars:
the dividing luminosity indicates
when the accretion disk changes mode of accretion, becoming radiatively
inefficient for luminosities of the order of 0.3--1 per cent of 
the Eddington one.

\item
The 10 BAT blazars at $z>2$ are among the most powerful known.
Not only their beamed jet bolometric luminosity, but also their
jet power and the accretion luminosities are among the largest ones.

\item 
The black hole masses are also very large, to account for the observed
disk luminosities and spectra.
They are all greater than one billion solar masses and a few approach
10 billions.

\item The ``record holder" S5 0014+813, having a black hole mass of 40 
billion solar masses, is an outlier with respect to the jet power -- disk
luminosity correlation defined by the {\it Fermi}/LAT FSRQs and obeyed by the 
high redshift BAT blazars.
This leads us to favour the hypothesis that its accretion disk is non--standard,
having, in its inner regions, a funnel collimating the 
radiation around the jet axis.
An anisotropy factor $\sim$10 is enough to make this object consistent with the jet power --
disk luminosity correlation.
Consequently, its  mass could be smaller (by a similar factor 10).

\item For each blazar pointing at us and detected through its beamed 
non--thermal emission, there should be other $\sim \Gamma^2$
at the same redshift with similar properties, including the mass of the black hole.
This puts a  lower limit to the density
of heavy black holes in the $3<z<4$ redshift range.

\item
Hard X--ray surveys can catch powerful and distant blazars where their high
energy SED peaks.
This implies that future X--ray missions such as 
{\it EXIST} and  {\it NHXM}  will be
the most effective way to find and study the most extreme radio--loud objects.
In the {\it Fermi}/LAT 0.1--100 GeV energy range these objects, having
steep spectra, are less conspicuous and can be missed by {\it Fermi}.
According to the {\it EXIST} sensitivity in hard X--rays,
powerful FSRQs can be easily detected even at redshift 8, if they exist.
According to the X--ray luminosity function derived by Ajello et al. (2009),
{\it EXIST} should detect 500--1500 blazars at $z>4$ and 20--60 at $z>6$,
allowing us to derive the mass function of 
radio--loud AGN up to very large redshifts.

\end{itemize}

\section*{Acknowledgments}
We thank the anonymous referee for his/her comments.
We thank Andrea Merloni and Marco Ajello for discussions.
This work was partly financially supported by a 2007 COFIN-MIUR 
and an ASI I/088/06/0) grants.
This research made use of the NASA/IPAC Extragalactic Database (NED) 
which is operated by the Jet Propulsion Laboratory, Caltech, under contract 
with NASA, and of the {\it Swift} public data
made available by the HEASARC archive system.
We also thank Neil Gehrels and the {\it Swift} team for quickly 
approving and performing the requested ToO observations of 1210+330.

\end{document}